\documentclass[11pt, a4paper]{article}
\usepackage[utf8]{inputenc}
\usepackage{amsmath,setspace, geometry}
\usepackage{amsthm}
\usepackage{amsfonts}
\usepackage{mathtools}
\mathtoolsset{showonlyrefs}
\usepackage[shortlabels]{enumitem}
\usepackage{rotating}
\usepackage{pdflscape}
\usepackage{graphicx}
\usepackage{bbm}
\usepackage[dvipsnames]{xcolor}
\usepackage[colorlinks=true, linkcolor= RawSienna, citecolor = RawSienna, filecolor = RawSienna, urlcolor = RawSienna, hypertexnames = true, backref = page]{hyperref}
\usepackage[]{natbib} 
\bibpunct[:]{(}{)}{,}{a}{}{,}
\usepackage[english]{babel}
\usepackage{float}
\usepackage{caption}
\usepackage{subcaption}
\usepackage{tikz}
\usepackage{booktabs}
\usepackage{pdfpages}
\usepackage{threeparttable}
\usepackage{framed}
\usepackage{comment}
\usepackage{lscape}
\usepackage{bm}
\usepackage[T1]{fontenc}
\usepackage{mlmodern}  
\DeclareFontFamily{OMX}{mlmex}{}
\DeclareFontShape{OMX}{mlmex}{m}{n}{<->mlmex10}{} 
\usepackage{tgtermes} 

\newtheorem{theorem}{Theorem}
\newtheorem{assumption}{Assumption}
\newtheorem{lemma}{Lemma}
\newtheorem{definition}{Definition}

\newtheorem{claim}{Claim}
\newtheorem{corollary}{Corollary}

\newtheorem{remark}{Remark}

\newtheorem*{theorem*}{Theorem}
\newtheorem*{assumption*}{Assumption}
\newtheorem*{lemma*}{Lemma}
\newtheorem*{definition*}{Definition}
\newtheorem*{proposition*}{Proposition}
\newtheorem*{claim*}{Claim}
\newtheorem*{corollary*}{Corollary}
\newtheorem*{example*}{Example}
\newtheorem*{result*}{Result}
\newtheorem*{remark*}{Remark}

\makeatletter
\def\@fnsymbol#1{} 
\makeatother

\makeatletter
\newcommand{\appendixsection}{%
  \setcounter{section}{0}%
  \renewcommand{\thesection}{\Alph{section}}%
  \renewcommand{\thetheorem}{\Alph{section}.\arabic{theorem}}%
  \renewcommand{\thelemma}{\Alph{section}.\arabic{lemma}}%
  \renewcommand{\theproposition}{\Alph{section}.\arabic{proposition}}%
  \renewcommand{\thecorollary}{\Alph{section}.\arabic{corollary}}%
  \renewcommand{\thedefinition}{\Alph{section}.\arabic{definition}}%
}
\makeatother

\title{Revisiting the Identification of the Conduct Parameter in Homogeneous Goods Markets\thanks{This paper is based on Chapter 3 of Matsumura's dissertation submitted to Rice University and is an extension of a paper circulated as "Identifying Conduct Parameters with Separable Demand: A Counterexample to Lau (1982)". 
We thank Jeremy Fox for his valuable advice.
We also thank the participants of Professor Fox's group meeting for their helpful comments.
This project was supported by JST ERATO Grant Number JPMJER2301, and Otani was supported by JSPS Grants 24K22604 and 25K16620.
The content of this paper is the authors' personal opinions and does not necessarily reflect the views of the Japan Fair Trade Commission.\\
Matsumura (corresponding author): The Japan Fair Trade Commission, \href{mailto:yuri.matsumura23@gmail.com}{yuri.matsumura23@gmail.com}.\\
Otani: Market Design Center, Department of Economics, University of Tokyo, \href{mailto:suguru.otani@e.u-tokyo.ac.jp}{suguru.otani@e.u-tokyo.ac.jp}
}}
\author{Yuri Matsumura \and Suguru Otani
}

\begin{document}
\maketitle
\begin{abstract}
    We revisit the identification of the conduct parameter in homogeneous goods markets.
    \citet{lauIdentifying1982} shows that the conduct parameter is not identified if and only if the inverse demand function is separable, except for a specific separable function.
    This result has been regarded as an extension of the result in \citet{bresnahanOligopoly1982} to more general settings.
    However, we show that Lau’s claim is incorrect and provide a new characterization of the non-identification.
    Our characterization shows that the presence of demand rotation instruments in the demand function is the necessary and sufficient condition for identifying the conduct parameter.
    Therefore, our result properly generalizes the role of demand rotation instruments in identifying the conduct parameter, as highlighted by \citet{bresnahanOligopoly1982}, to more general settings.
\end{abstract}

\noindent\textbf{Keywords:} Identification, Conduct Parameter, Homogeneous Goods Market
\vspace{0in}
\newline
\noindent\textbf{JEL Codes:} C5, C13, L1

\newpage
\section{Introduction}


Empirical research in industrial organization examines a wide range of questions, including the measurement of market power and concentration, the evaluation of welfare, the assessment of cartels, and the prediction of merger impacts.
In virtually all of these applications, empirical conclusions depend critically on assumptions about how firms compete, that is, on the specification of firm conduct.
Researchers often impose a particular competitive model, such as price-taking, Cournot, or Bertrand, because conduct is rarely identified without additional information, such as detailed cost data or industry- or institution-specific information. 
Nevertheless, misspecifying conduct can distort the measurement or lead to misleading conclusions.

To address this problem, the \textit{conduct parameter approach} has been used in the literature on industrial organization.
This approach embeds a scalar measure of firm conduct into the marginal revenue function and allows researchers to estimate conduct directly from data, providing a flexible model to investigate several questions.
The applications vary across industries and questions, including homogeneous-product markets \citep*{porterStudy1983, genesoveTesting1998, okazakiExcess2022}, differentiated-product markets \citep*{millerUnderstanding2017, cilibertoDoes2014, sullivanIce2020}, pass-through analysis \citep*{weylPassThrough2013, millerPassthrough2017}, welfare evaluation under imperfect competition \citep*{beringeRobust2025, nocke2025concentration}, and merger evaluation \citep*{aryalBridging2025}.

Despite its broad use, the basic identification result on the conduct parameter in a homogeneous goods market remains rooted in two classic contributions.
First, \citet{bresnahanOligopoly1982} shows that, with a linear demand and a linear marginal cost, a \textit{demand rotation instrument} can identify the conduct parameter by moving the slope and intercept of inverse demand simultaneously because
it can keep the equilibrium the same under the true conduct parameter while changing the equilibrium under the false conduct parameter.
Second, \citet{lauIdentifying1982} considers an environment with a more general inverse demand function and a marginal cost function and shows that conduct is not identified if and only if the inverse demand function is separable in demand shifters.
Because demand rotation instruments break the separability of the inverse demand function, Lau's condition has been interpreted as a general characterization of identification and regarded as an extension of Bresnahan's idea in general settings.

The main contribution of this paper is to show that Lau's characterization is incomplete and to provide a new condition that fully characterizes if and only if conduct and marginal cost are not identified.
Our main result shows that the non-identification of the conduct parameter and the marginal cost function arises if and only if demand shifters can change \textit{only the slope} or \textit{only the intercept} of the inverse demand function.
In this case, the inverse demand function cannot have any demand rotation instrument, and hence conduct is identified precisely when the inverse demand function includes demand rotation instruments.
This result is both simpler and more restrictive than separability in Lau's claim and clarifies the precise role of demand rotation instruments in identifying the conduct parameter in general environments.

The intuition is the following.
Given a conduct parameter and a marginal cost function, the non-identification implies that we can transform the marginal cost function into another marginal cost function that, together with another conduct parameter, produces the same equilibrium quantity.
However, this transformation is usually invalid because keeping the equilibrium points the same in both models requires the transformed marginal cost function to depend on demand shifters, which violates the exclusive demand shifters assumption.
This dependence disappears only when demand shifters can change just the slope or just the intercept of the inverse demand function, in which case we can construct different pairs of conduct parameters and marginal cost functions that yield the same equilibrium for any value of the shifters without violating the exclusion restriction. 
Thus, when demand rotation instruments are present in the inverse demand function, the transformation remains invalid, preventing observational equivalence between alternative conduct and marginal-cost pairs and ensuring identification.
The issue that makes Lau's proof incomplete is that while he also derives a transformation between marginal cost functions, he fails to eliminate the dependence of the transformed marginal cost function on demand shifters.

Taken together, our results refine the theoretical foundations of conduct parameter estimation and clarify the exact role that demand shifters must play to allow practitioners to implement empirical analysis with a flexible competition model. 
The rest of the paper is organized as follows.
Section \ref{sec:setting} describes the setting.
Section \ref{sec:main_result_and_lau_result} discusses Lau's result and our main result.
Subsections \ref{subsec:necessary_condition_nonidentification} and \ref{subsec:sufficient_condition_nonidentification} provide a new characterization of non-identification and show the necessary and sufficient condition for non-identification.
Section \ref{sec:other_discussions} discusses some criticisms of the conduct parameter approach.
Section \ref{sec:conclusion} concludes.
The Appendix includes the omitted proofs in the main text and a summary of \citet{goldmanNote1964}.




\section{Setting}\label{sec:setting}

\subsection{Conduct Parameter Model}\label{subsec:model_and_assumptions}

Consider a homogeneous product market with an aggregate inverse demand and an aggregate marginal cost function denoted by $P(Q, \bm{X}^{d})$ and $MC(Q, \bm{X}^{s})$, respectively, where $Q$ is the aggregate product quantity, $\bm{X}^{d}$ and $\bm{X}^{s}$ are the vectors of demand shifters and cost shifters, respectively.
Note that every vector is represented by a bold letter, and its element or scalar value is represented by a non-bold letter.
Let $K_d$ and $K_s$ be the dimension of $\bm{X}^{d}$ and $\bm{X}^{s}$, respectively.

Given the demand shifter $\bm{X}^{d}$ and the cost shifter $\bm{X}^{s}$, the equilibrium quantity $Q^e$ solves the following equation:
\begin{align}
    P(Q, \bm{X}^{d}) + \theta Q\frac{\partial P}{\partial Q}(Q, \bm{X}^{d}) = MC(Q, \bm{X}^{s}), \label{eq:foc}
\end{align}
where $\theta\in [0,1]$ is called the conduct parameter.
By rewriting the equilibrium condition \eqref{eq:foc}, we have 
\begin{align}
    \theta = \frac{P - MC}{P}\varepsilon^{d},
\end{align}
where $\varepsilon^{d}$ is the price elasticity of demand.
Therefore, the conduct parameter is also regarded as the elasticity-adjusted Lerner index and represents the degree of market power of the firms.
Depending on the value of $\theta$, the condition can represent the first-order condition of several models between perfect competition ($\theta=0$) and joint-profit maximization ($\theta=1$).\footnote{It can also nest Cournot competition when $\theta=1/N$ under some marginal cost function such as constant marginal cost and linear marginal cost. In general, this holds when the aggregation of each firm's first-order condition results in the aggregate first-order condition.}
Therefore, the left-hand side of \eqref{eq:foc} is regarded as a generalized marginal revenue with $\theta$, and hence the condition is a generalized first-order condition with the conduct parameter.

For the identification analysis, we put some restrictions.
First, we assume that the demand shifter and the cost shifter are mutually exclusive:
\begin{assumption}\label{assumption:exclusive_shifters}
    The set of all exogenous variables affecting the market equilibrium can be partitioned into (1) exclusive demand shifters, $\bm{X}^{d}$, that affect the inverse demand function  but not the marginal cost function, (2) exclusive cost shifters, $\bm{X}^{s}$, that affect the marginal cost function but not the inverse demand function, and (3) common shifters, $\bm{Z}$, that affect both the inverse demand function and the marginal cost function.
\end{assumption}

The theoretical analysis of identification often simplifies the setting by assuming that the demand shifters and cost shifters are mutually exclusive, as per Assumption \ref{assumption:exclusive_shifters}. 
We formalize our approach by conditioning the analysis on an arbitrary realization $\bm{z}$ of these common shifters. This means all subsequent functions are understood to be conditional functions, $P(Q, \bm{X}^d) \equiv \tilde{P}(Q, \bm{X}^d, \bm{Z} = \bm{z})$ and $MC(Q, \bm{X}^s) \equiv \tilde{MC}(Q, \bm{X}^s, \bm{Z} = \bm{z})$. By treating $\bm{Z}$ as fixed constants in this conditional analysis, we maintain the notational simplicity of focusing only on the exclusive shifters while ensuring that the derived non-identification characterization holds robustly across all values of the common variables in empirical settings.

The next assumption restricts the effectiveness of the shifters:
\begin{assumption}\label{assumption:effectiveness_shifters}
    The given demand shifters and the cost shifters should affect the inverse demand function and the marginal cost function.
    Formally, we should have that for all $i = 1, \ldots, K_d$ and $j = 1, \ldots, K_s$,
    \begin{align}
        \frac{\partial P}{\partial X^{d}_i}(Q, \bm{X}^{d}) \ne 0 \quad \text{and} \quad \frac{\partial MC}{\partial X^{s}_j}(Q, \bm{X}^{s}) \ne 0
    \end{align}
    for any $Q>0$, $\bm{X}^{d}$, and $\bm{X}^{s}$. 
\end{assumption}

When this condition is not met, there could be areas where the inverse demand function and the marginal cost function cannot be identified.
For example, when a demand shifter does not change the inverse demand function for some interval of the demand shifter, the equilibrium quantity is not affected by the change in the demand shifter.
Then, there is no variation to identify the inverse demand function on the interval.
Because our identification result for the conduct parameter and the marginal cost function presumes the identification of the inverse demand function, to guarantee the identification of the inverse demand function, the assumption is necessary.

Next, we impose an assumption on the differentiability of the inverse demand function and the marginal cost function:
\begin{assumption}\label{assumption:three_times_differentiable}
    The inverse demand function is three times continuously differentiable, and the marginal cost function is twice continuously differentiable.
\end{assumption}
As we will see, \citet{lauIdentifying1982} only assumes twice-continuous differentiability of the inverse demand function and the marginal cost function.
Hence, our result requires a stronger assumption on the inverse demand function.
However, we will justify our assumption by showing that Lau's claim also needs three-times continuous differentiability later.

Finally, we impose the equilibrium existence condition:
\begin{assumption}\label{assumption:unique_equilibrium}
    Given an inverse demand function, a conduct parameter, and a marginal cost function, and given $\bm{X}^{d}$ and $\bm{X}^{s}$, the derivative of the equilibrium condition with respect to $Q$ at the equilibrium quantity $Q^{e}$ is not zero, that is,
    \begin{align}
        (1+\theta)\frac{\partial P}{\partial Q}(Q^{e}, \bm{X}^{d}) + \theta Q^{e}\frac{\partial^2 P}{\partial Q^2}(Q^{e}, \bm{X}^{d}) - \frac{\partial MC}{\partial Q}(Q^{e}, \bm{X}^{s}) & \ne 0.
    \end{align}
\end{assumption}
Note that as the above equation consists of the derivative of the inverse demand and the marginal cost, by Assumption \ref{assumption:three_times_differentiable}, the derivative of the equilibrium condition with respect to $Q$ is finite, and hence the left-hand side is a finite value for any $Q$, $\bm{X}^{d}$, and $\bm{X}^{s}$.

\subsection{The Data Generation Process}\label{subsec:data_generation_process}

Suppose that the researcher observes the aggregate price $P$ and the aggregate quantity $Q$, and the vector of exogenous variables $\bm{X}^{d}$ and $\bm{X}^{s}$.
Assume that the data is generated through the equilibrium condition \eqref{eq:foc}.
We assume that from the data, the reduced form of the equilibrium quantity and the equilibrium price are identified:
\begin{assumption}\label{assumption:reduced_form}
    The reduced form of the equilibrium quantity and the equilibrium price, defined as
    \begin{align}
        Q^e = h_q(\bm{X}^{d}, \bm{X}^{s}), \quad P^e = h_p(\bm{X}^{d}, \bm{X}^{s}),
    \end{align}
    are identified.
\end{assumption}
The identification of the reduced forms follows directly from variation in the demand and cost shifters.

\begin{remark}
    While \citet{bresnahanOligopoly1982} considers a model with error terms, by following \citet{lauIdentifying1982}, we do not consider any error term in the demand and the supply side, that is, our model does not have any unobserved characteristics to the econometrician.
    This is a restrictive assumption because there could be real data where $\bm{X}^d$ and $\bm{X}^s$ are the same in two time periods, but the equilibrium outcomes are different in the two periods.
    As the scope of this paper is to fully characterize the identification condition along with Lau's setting,. In other words, our identification analysis is independent of the presence of error terms and concerns whether the structural mapping from primitives $(\theta, MC)$ to the reduced-form equilibrium outcomes $(h_q, h_p)$ is injective, that is, whether $(\theta, MC)$ can be uniquely recovered given $(h_q, h_p)$. We do not include unobserved characteristics and leave it for future research.
\end{remark}

\subsection{Definitions}\label{subsec:definitions}
Now, we introduce the definition of the identification problem in this setting.
While our interest is the identification of the conduct parameter and the marginal cost function, by following \citet{lauIdentifying1982}, we take an indirect approach and specify the conditions under which the model is \textit{not} identified.
Our definition of non-identification is based on when two different models lead to observational equivalence:
\begin{definition}\label{definition:non_identification}
    The conduct parameter and the marginal cost function are said to be non-identified if there are two distinct pairs of conduct parameters and marginal cost functions, denoted by $(\theta, MC)$ and $(\theta^{*}, MC^{*})$, such that the corresponding reduced-form equilibrium quantity that satisfies the equilibrium condition \eqref{eq:foc} and the corresponding reduced-form equilibrium price are identical for every demand- and cost-shifter value: $$Q^{e} = h_q(\bm{X}^{d}, \bm{X}^{s}) = h_q^{*}(\bm{X}^{d}, \bm{X}^{s}) \text{ and } P^{e} = h_p(\bm{X}^{d}, \bm{X}^{s}) = h_p^{*}(\bm{X}^{d}, \bm{X}^{s})$$ for any $\bm{X}^{d}$ and $\bm{X}^{s}$. 
\end{definition}

By taking the contraposition of Definition \ref{definition:non_identification}, we can characterize the identification of the conduct parameter and the marginal cost function:
\begin{corollary}\label{corollary:identification}
    The conduct parameter and the marginal cost are said to be identified if for any two distinct pairs of a conduct parameter and a marginal cost function, there exists a pair of demand shifters $\tilde{\bm{X}}^{d}$ and $\tilde{\bm{X}}^{s}$ where the equilibrium quantity $Q^e$ is different, that is, $h_q(\tilde{\bm{X}}^{d}, \tilde{\bm{X}}^{s}) \ne h_q^{*}(\tilde{\bm{X}}^{d}, \tilde{\bm{X}}^{s})$.
\end{corollary}
Note that the non-identification and the identification assume that the inverse demand function is already identified.
In other words, we study the identification of the conduct parameter and the marginal cost function given an identified inverse demand function.

\section{Main Result: Characterization of the Non-identification}\label{sec:main_result_and_lau_result}
\subsection{Lau's Claim}
\citet{lauIdentifying1982} investigates the identification of the conduct parameter and the marginal cost function and obtains the following claim, which is the quote of Theorem 1 in \citet{lauIdentifying1982}:
\begin{claim}\label{theorem_lau}
    Under the assumption that the inverse demand function and the marginal cost functions are twice continuously differentiable, the index of competitiveness $\theta$ cannot be identified from data on industry price and quantity and other exogenous variables alone if and only if the industry inverse demand function is separable in the demand shifter, that is,
    \begin{align}
        P(Q,\bm{X}^{d}) = P(Q, r(\bm{X}^{d})), \label{eq:separable_demand}
    \end{align}
    but does not take the form 
    \begin{align}
        P(Q, r(\bm{X}^{d})) = Q^{-\frac{1}{\theta}}r(\bm{X}^{d}) + s(Q). \label{eq:identification_separable_demand_lau}
    \end{align}
\end{claim}

The result states that the separability of the inverse demand function is crucial for the identification, but there is a type of separable inverse demand function that can lead to the identification.
An inverse demand function satisfying \eqref{eq:separable_demand} has weak separability defined in \citet{goldmanNote1964}.\footnote{See Definition \ref{def:weak_separable} in Appendix \ref{appendix:summary_goldman_uzawa}.
Note that the separability in \citet{goldmanNote1964} is defined only when the dimension of the demand shifter is greater than two. When the demand shifter is a scalar, we cannot apply the definition of separability in \citet{goldmanNote1964}.}
The claim also emphasizes that the dimension of the demand shifter is important for the identification.
When the demand shifter is a scalar, \eqref{eq:separable_demand} nests any inverse demand function because we can set $\tilde{P}(Q, \bm{X}^{d}) = P(Q, r(\bm{X}^{d}))$.
Thus, with a scalar demand shifter, the conduct parameter can be identified only when the inverse demand function takes the form \eqref{eq:identification_separable_demand_lau}.

To understand the intuition of Lau's claim, let us consider an example in \citet{bresnahanOligopoly1982} without an error term.
Bresnahan considers a market with a linear inverse demand and a linear marginal cost, where the linear inverse demand function is given as
\begin{align}
    P(Q, \bm{X}^{d}) = \alpha_0 - \alpha_1 Q + \alpha_2X^{d}_1 + \alpha_3QX^{d}_1 + \alpha_4X^{d}_2, \label{eq:demand_bresnahan}
\end{align}
where $X^{d}_1$ is a demand rotation instrument because it can change the slope and the intercept of the inverse demand function without changing the equilibrium quantity.
It is easy to verify that the demand rotation instrument breaks the separability of the inverse demand function.
Additionally, the dimension of the demand shifter is greater than one, and hence Lau's claim implies that even a nonlinear marginal cost function, \eqref{eq:demand_bresnahan}, leads to the identification of the conduct parameter.

While Lau's claim has been regarded as an extension of \citet{bresnahanOligopoly1982} to more general settings, an important observation is that including demand-rotation instruments is inconsistent with imposing separability on the inverse demand function, but the separability assumption itself does not preclude the existence of demand-rotation shifters in an inverse demand function.

For example, consider an inverse demand function given as 
\begin{align}
    P(Q, r(\bm{X}^d)) = -Qr(\bm{X}^{d}) + Q + r(\bm{X}^{d}) = r(\bm{X}^{d})(1- Q) - Q, 
\end{align}
where the range of $r$ is greater than one, that is, $r(\bm{X}^{d})\ge 1$ for any $\bm{X}^{d}$.
Assume that the dimension of $\bm{X}^d$ is greater than two.
Note that in this inverse demand function, the demand shifter can change the slope and the intercept of the inverse demand function.

First, the function is verified to be separable because for any $i$ and $j$, we have
\begin{align}
    \frac{\partial}{\partial Q}\left(\frac{\frac{\partial P}{\partial X^{d}_i}(Q,\bm{X}^d)}{\frac{\partial P}{\partial X^{d}_j}(Q,\bm{X}^d)}\right) = \frac{\partial}{\partial Q}\left(\frac{r_i(\bm{X}^{d})(1- Q)}{r_j(\bm{X}^{d})(1- Q)}\right) = \frac{\partial}{\partial Q}\left(\frac{r_i(\bm{X}^{d})}{r_j(\bm{X}^{d})}\right) = 0,
\end{align}
where $r_i(\bm{X}^d) \equiv \frac{\partial r(\bm{X}^{d})}{\partial \bm{X}^{d}_i}$.

Second, we can verify that the demand shifters can work as demand rotation instruments to break observational equivalence.
Given the inverse demand function, the marginal revenue under $\theta$ is written as
\begin{align}
    MR(Q, X^{d};\theta) = -r(\bm{X}^{d})(1 - (1+\theta)Q) + Q(1+\theta).
\end{align}
When the quantity is $Q' = \frac{1}{1+\theta}$, the marginal revenue is not affected by the change in the demand shifter $\bm{X}^{d}$ because it is equal to one.
Suppose that there is a marginal cost function $MC$ where $Q'$ holds as an equilibrium for some $\tilde{\bm{X}^{s}}$.
Then, as the marginal revenue is a constant in the demand shifter, any change in the demand shifter does not affect the equilibrium quantity.
In contrast, suppose that under another conduct parameter $\theta^{*}$ and another marginal cost function $MC^{*}$, $Q'$ holds as an equilibrium for some $\tilde{\bm{X}^{d}}$ and $\tilde{\bm{X}^{s}}$ , that is, $Q'$ satisfies
\begin{align}
    r(\tilde{\bm{X}^{d}})(1 - (1+\theta^{*})Q') + Q'(1+\theta^{*}) = MC^{*}(Q', \tilde{\bm{X}^{s}}).
\end{align}
Here, the marginal revenue under $\theta^{*}$ is a function of the demand shifter at $Q'$.
Thus, the change in $\bm{X}^{d}$ leads to a different equilibrium quantity from $Q'$, which implies the violation of the observational equivalence.
Therefore, the demand shifter $\bm{X}^{d}$ can work as the demand rotation instrument at $Q'$ because it changes the slope and the intercept of the inverse demand function simultaneously without changing the equilibrium quantity under the true conduct, whereas the change leads to a different equilibrium quantity under any false conduct.

Furthermore, when we treat $r(\bm{X}^{d})$ as a scalar demand shifter, the second argument implies that even a scalar demand shifter can break observational equivalence, although the inverse demand function does not satisfy \eqref{eq:identification_separable_demand_lau}.
The example admits a separable inverse demand function with a demand rotation instrument, but Lau does not clearly explain how this situation fits his claim, and hence it is not clear why separability, rather than the availability of demand rotation instruments, is the key to identification.

\subsection{Main Result}
We now show the following theorem:
\begin{theorem}\label{theorem:identification_characterization}
    Given Assumption \ref{assumption:exclusive_shifters}, \ref{assumption:effectiveness_shifters}, \ref{assumption:three_times_differentiable}, and  \ref{assumption:unique_equilibrium}, and data on price, quantity, and other exogenous variables, 
    \begin{itemize}
        \item[(i)] if the conduct parameter $\theta$ and the marginal cost function $MC$ are not identified, the industry inverse demand function must be given as
        \begin{align}
            P(Q, \bm{X}^{d}) = Q^{\alpha}r(\bm{X}^{d}) + s(Q) \label{eq:nonidentification_inverse_demand}
        \end{align}
        where $\alpha \ne -\frac{1}{\theta}$.
        \item[(ii)] if the inverse demand function takes the form \eqref{eq:nonidentification_inverse_demand} with the properties described in (i), then the conduct parameter and the marginal cost are not identified.
    \end{itemize}
\end{theorem}
As in Lau's claim, when the demand shifter is a vector, the inverse demand function is separable because the demand shifters affect the inverse demand function only through $r$.
However, our result tells more about under what type of separable function the non-identification holds and has a clear economic interpretation.
As in Claim \ref{theorem_lau}, our result has a special case where any change in demand shifters does not affect the marginal revenue under the true conduct but does under any false conduct, and hence the identification always holds ($\alpha = - \frac{1}{\theta}$).
Except in this case, observe that the demand shifter changes only the slope of the inverse demand function when $\alpha \ne 0$ or changes only the intercept of the inverse demand function when $\alpha = 0$.
Recall that demand rotation instruments can change the intercept and the slope of the inverse demand function simultaneously.
Therefore, we cannot have any demand rotation instruments under \eqref{eq:nonidentification_inverse_demand}.
The result implies that the identification holds if and only if the inverse demand function has demand shifters that work as demand rotation instruments.
Here, we are explicit that a demand rotation instrument is not necessarily a single variable, but could be a combination of changes in some demand shifters that alter the slope and the intercept of the inverse demand function simultaneously.
Note also that Bresnahan graphically demonstrates that the conduct parameter is identified when a demand rotation instrument moves the demand curve in a way that leaves the equilibrium point unchanged at the true conduct but changes the equilibrium under any false conduct.
However, this is an extreme case and it suffices that the instrument changes the slope and the intercept simultaneously.
This clearly reinforces the idea of \citet{bresnahanOligopoly1982} using demand rotation instruments to identify the conduct parameter in more general settings.

Hereafter, we provide the proof of Theorem \ref{theorem:identification_characterization}.
First, we derive the necessary condition for non-identification based on Definition \ref{definition:non_identification}.
By rewriting Definition \ref{definition:non_identification}, we obtain a transformation that maps the derivatives of $MC$ into the derivatives of another marginal cost function $MC^{*}$ that yields observational equivalence given an inverse demand function (Lemma \ref{lemma:mc_transformation}).
However, the transformation is not valid because it consists of derivatives of the inverse demand function and hence depends on the demand shifter $\bm{X}^{d}$.
Without additional restrictions, $MC^{*}$ would therefore depend on $\bm{X}^{d}$, violating Assumption \ref{assumption:exclusive_shifters}.
Therefore, to make the transformation valid, we need to remove the effect of the demand shifter from the transformation, which puts restrictions on the inverse demand function.
Then, the restriction leads to a differential equation that characterizes the inverse demand function that leads to non-identification (Lemma \ref{lemma:nonidentification_inverse_demand}).
Then, to show sufficiency, under the inverse demand function characterized in Lemma \ref{lemma:nonidentification_inverse_demand}, we construct a transformation that leads to observational equivalence (Lemma \ref{lemma:sufficient_nonidentification}).

\subsection{Necessary Condition for the Non-identification}\label{subsec:necessary_condition_nonidentification}

First, we characterize the non-identification condition based on Definition \ref{definition:non_identification}.
The characterization is based on the fact that observational equivalence implies that the reduced forms of the equilibrium quantity $h_q$ and $h_q^{*}$ are identical for any $\bm{X}^{d}$ and $\bm{X}^{s}$.
Therefore, its derivative with respect to the demand and cost shifters, $\nabla h_q$ and $\nabla h_q^{*}$, should also be identical.
Then, by applying the implicit function theorem to the equilibrium condition, we can compute $\nabla h_q$ and $\nabla h_q^{*}$.
The following lemma characterizes the non-identification condition:
\begin{lemma}\label{lemma:nonidentification_characterization}
    Non-identification implies that for any $\bm{X}^{d}$, $\bm{X}^{s}$, and $Q^e$ under these exogenous variables, we have for $i = 1, \ldots, K_d$,
    \begin{align}
        &\frac{\partial P}{\partial X^{d}_i}(Q^e, \bm{X}^{d}) + \theta^{*} Q^e \frac{\partial^2 P}{\partial X^{d}_i\partial Q}(Q^e, \bm{X}^{d})\\  
        &= \lambda(Q^e, \bm{X}^{d}, \bm{X}^{s})\left[ \frac{\partial P}{\partial X^{d}_i}(Q^e, \bm{X}^{d}) + \theta Q^e \frac{\partial^2 P}{\partial X^{d}_i\partial Q}(Q^e, \bm{X}^{d}) \right], \label{eq:nonidentification_demand}
    \end{align}
    and for $j = 1,\ldots, K_s$,
    \begin{align}
        \frac{\partial MC^{*}}{\partial X^{s}_j}(Q^e, \bm{X}^{s}) = \lambda(Q^e, \bm{X}^{d}, \bm{X}^{s}) \frac{\partial MC}{\partial X^{s}_j}(Q^e, \bm{X}^{s}),\label{eq:nonidentification_marginal_cost}
    \end{align}
    where $\lambda(Q^e, \bm{X}^{d}, \bm{X}^{s})$ is defined as
    \begin{align}
        \lambda(Q^e, \bm{X}^{d}, \bm{X}^{s}) \equiv \frac{(1+\theta^{*})\frac{\partial P}{\partial Q}(Q^e, \bm{X}^{d}) + \theta^{*} Q^e\frac{\partial^2 P}{\partial Q^2}(Q^e, \bm{X}^{d}) - \frac{\partial MC^{*}}{\partial Q}(Q^e, \bm{X}^{s})}{(1+\theta)\frac{\partial P}{\partial Q}(Q^e, \bm{X}^{d}) + \theta Q^e\frac{\partial^2 P}{\partial Q^2}(Q^e, \bm{X}^{d}) - \frac{\partial MC}{\partial Q}(Q^e, \bm{X}^{s})}. \label{eq:lambda_foc}
    \end{align}
\end{lemma}
See Appendix \ref{appendix:proof} for the detailed proof.
Equation \eqref{eq:nonidentification_demand} and \eqref{eq:nonidentification_marginal_cost} imply that when non-identification holds, the derivative of the marginal revenue with $\theta$ can be transformed into the derivative of the marginal revenue with $\theta^{*}$, and the derivative of the marginal cost $MC$ with respect to $\bm{X}^{s}$ can be transformed into the derivative of the marginal cost $MC^{*}$ with respect to $\bm{X}^{s}$ by $\lambda(Q^e, \bm{X}^{d}, \bm{X}^{s})$.

These transformations cannot always be valid because the transformed marginal revenue is affected by the cost shifter $\bm{X}^{s}$ and the transformed marginal cost is affected by the demand shifter $\bm{X}^{d}$ through $\lambda$.
These dependencies are not allowed by Assumption \ref{assumption:exclusive_shifters}.
Thus, to consider a valid transformation, we further rewrite \eqref{eq:nonidentification_demand} and \eqref{eq:nonidentification_marginal_cost}.
The next lemma provides a transformation of the derivative of $MC$ and $MC^{*}$ with respect to $Q$ and $\bm{X}^{s}$ that leads to observational equivalence:
\begin{lemma}\label{lemma:mc_transformation}
    For any $Q^e$, $\bm{X}^{d}$, and $\bm{X}^{s}$, non-identification implies that the derivative of marginal cost $MC$ can be transformed into the derivative of marginal cost $MC^{*}$: for $i = 1, \ldots, K_d$, and $j = 1, \ldots, K_s$,
    \begin{align}
        \frac{\partial MC^{*}}{\partial Q}(Q^e, \bm{X}^{s}) =D_i(Q^e, \bm{X}^{d}) + C_i(Q^e, \bm{X}^{d})\frac{\partial MC}{\partial Q}(Q^e, \bm{X}^{s}),\label{eq:mc_transformation_quantity}
    \end{align}
    and
    \begin{align}
        \frac{\partial MC^{*}}{\partial X^{s}_j}(Q^e, \bm{X}^{s}) = C_i(Q^e, \bm{X}^{d})\frac{\partial MC}{\partial X^{s}_j}(Q^e, \bm{X}^{s}).\label{eq:mc_transformation_cost_shifter}
    \end{align}
    where $C_i(Q, \bm{X}^{d})$ and $D_i(Q, \bm{X}^{d})$ are defined as
    \begin{align}
        C_i(Q, \bm{X}^{d}) \equiv \frac{\frac{\partial P}{\partial X^{d}_i}(Q, \bm{X}^{d}) + \theta^{*} Q\frac{\partial^2 P}{\partial X^{d}_i\partial Q}(Q, \bm{X}^{d}) }{\frac{\partial P}{\partial X^{d}_i}(Q, \bm{X}^{d}) + \theta Q\frac{\partial^2 P}{\partial X^{d}_i\partial Q}(Q, \bm{X}^{d}) },\label{eq:ratio_marginal_revenue}
    \end{align}
    and
    \begin{align}
        D_i(Q, \bm{X}^{d}) & \equiv\frac{\theta^{*} - \theta}{\frac{\partial P}{\partial X^{d}_i}(Q, \bm{X}^{d}) + \theta Q\frac{\partial^2 P}{\partial X^{d}_i\partial Q}(Q, \bm{X}^{d})}\Bigg[\frac{\partial P}{\partial Q}(Q, \bm{X}^{d}) \frac{\partial P}{\partial X^{d}_i}(Q, \bm{X}^{d})\\
        &\quad + Q\frac{\partial^2 P}{\partial Q^2}(Q, \bm{X}^{d}) \frac{\partial P}{\partial X^{d}_i}(Q, \bm{X}^{d}) - Q \frac{\partial P}{\partial Q}(Q, \bm{X}^{d}) \frac{\partial^2 P}{\partial X^{d}_i\partial Q}(Q, \bm{X}^{d}) \Bigg]\label{eq:interaction_derivative_demand}
    \end{align}
\end{lemma}
See Appendix \ref{appendix:proof} for the detailed proof. 
The subscript $i$ in $C_i$ and $D_i$ indicates that $C_i$ and $D_i$ consist of the derivative of the inverse demand function with respect to $X^{d}_i$.
Unlike Lemma \ref{lemma:nonidentification_characterization}, we have only the transformation relating to marginal cost, and hence we can see the potential obstacle in this transformation more clearly: for the transformations to hold universally for any $Q^e$ and $\bm{X}^{s}$, they must be independent of $\bm{X}^{d}$.
If the dependence holds, while a demand rotation instrument does not change the equilibrium quantity $Q^e$, it changes $MC^{*}$ via $C_i$ and $D_i$.
However, this means that the marginal cost function is a function of the demand shifter $\bm{X}^{d}$, which is prohibited by Assumption \ref{assumption:exclusive_shifters}. 
Therefore, in order for the transformations to be valid, both $C_i$ and $D_i$ must be independent of $\bm{X}^{d}$.

To see when the terms $C_i$ and $D_i$ are independent of $\bm{X}^{d}$, we take the derivative of $C_i$ and $D_i$ with respect to $X^{d}_j$ and set them equal to zero for all $j = 1,\ldots, K_{d}$.
A concern is that, as $C_i$ and $D_i$ already consist of the second-order derivative of the inverse demand function, taking the derivative of $C_i$ and $D_i$ leads to three-times derivative of the inverse demand function.
This is why we need a stronger restriction (Assumption \ref{assumption:three_times_differentiable}) on the inverse demand function than Lau considers (the twice-continuous differentiability assumption in Lau's claim).

\paragraph{Case: The denominators of $C_i$ and $D_i$ are zero for some demand shifters}
Before we characterize the inverse demand function that makes $C_i$ and $D_i$ independent of $\bm{X}^{d}$, we first check when $C_i$ and $D_i$ are not well-defined.
This implies that \eqref{eq:ratio_marginal_revenue} and \eqref{eq:interaction_derivative_demand} are violated, and hence the identification holds by Corollary \ref{corollary:identification}.
All derivatives in $C_i$ and $D_i$ are finite by Assumption \ref{assumption:three_times_differentiable}, and hence the numerator and the denominator of $C_i$ and $D_i$ cannot take an infinite value.
In this case, regardless of the numerator of $C_i$ or $D_i$, \eqref{eq:ratio_marginal_revenue} and \eqref{eq:interaction_derivative_demand} are violated whenever at least one denominator for $i= 1, \ldots, K_d$ is zero.
In this case, $C_i$ and $D_i$ will be infinite or indeterminate.

When the denominator of $C_i$ or $D_i$ is zero for some $i$, we have that for some $(\tilde{Q}, \tilde{\bm{X}}^{d})$,
\begin{align}
    \frac{\partial P}{\partial X^{d}_i}(\tilde{Q}, \tilde{\bm{X}}^{d}) + \theta \tilde{Q}\frac{\partial^2 P}{\partial X^{d}_i\partial Q}(\tilde{Q}, \tilde{\bm{X}}^{d}) = 0. \label{eq:identification_condition_separable}
\end{align}
Let $\mathcal{I}$ be the set of indices of the demand shifters where \eqref{eq:identification_condition_separable} holds.
Note that \eqref{eq:identification_condition_separable} can be rewritten as
\begin{align}
    \frac{\partial }{\partial X^{d}_i}\left( P(\tilde{Q}, \tilde{\bm{X}}^{d}) + \theta \tilde{Q} \frac{\partial P}{\partial Q}(\tilde{Q}, \tilde{\bm{X}}^{d})\right) = 0,
\end{align}
which is the derivative of the marginal revenue under $\theta$ with respect to $X^{d}_i$.
Therefore, this implies that the marginal revenue under $\theta$ is not affected by the change in $X^{d}_i$ at $(\tilde{Q}, \tilde{\bm{X}}^{d})$.
This also implies that the equilibrium quantity is not affected by the change in $\tilde{X}^{d}_i$ at $(\tilde{Q}, \tilde{\bm{X}}^{d})$.

The non-identification implies that \eqref{eq:identification_condition_separable} does not hold for any $Q^{e}$ and $\bm{X}^{d}$.
Once we have $(\tilde{Q}, \tilde{\bm{X}}^{d})$ for some $i$ where \eqref{eq:identification_condition_separable} holds, it suffices for identification.
While it is hard to characterize the inverse demand function that satisfies \eqref{eq:identification_condition_separable} for some $(\tilde{Q}, \tilde{\bm{X}}^{d})$ and for some $i$, we can characterize the inverse demand function that satisfies \eqref{eq:identification_condition_separable} for all $(Q, \bm{X}^{d})$.
Note that \eqref{eq:identification_condition_separable} is a partial differential equation that can be solved analytically.
Thus, we can characterize the inverse demand function that always leads to identification for any $Q$ and $\bm{X}^{d}$:
\begin{lemma}\label{lemma:identification_condition_separable}
    Suppose that $\theta \ne 0$ and $Q > 0$. Then, the conduct parameter and the marginal cost function are identified when the inverse demand function is such that
    \begin{align}
        P(Q, \bm{X}^{d}) = Q^{-\frac{1}{\theta}}r(\bm{X}^{d}) + s(Q, \bm{X}^{d}_{-\mathcal{I}}), \label{eq:identification_separable_demand}
    \end{align}
    where $\bm{X}^{d}_{-\mathcal{I}}$ is the vector of demand shifters whose index is not in $\mathcal{I}$, and $r: \mathbb{R}^{K_d} \rightarrow \mathbb{R}$ and $s: \mathbb{R}^{|\bm{X}^{d}_{-\mathcal{I}}|+1}$ are at least twice-continuously differentiable.
\end{lemma}
See Appendix \ref{appendix:proof} for the proof.
Under \eqref{eq:identification_separable_demand}, the equilibrium condition \eqref{eq:foc_theta} is given as
\begin{align}
    s(Q, \bm{X}^{d}_{-\mathcal{I}}) + \theta Qs'(Q, \bm{X}^{d}_{-\mathcal{I}}) = MC(Q, \bm{X}^{s}),
\end{align}
Therefore, when \eqref{eq:identification_condition_separable} holds, the equilibrium quantity is not affected by the change in $X^{d}_i$.
On the other hand, for any other $\theta^{*} \ne \theta$, the equilibrium condition becomes
\begin{align}
    Q^{-\frac{1}{\theta}}r(\bm{X}^{d})\left(1 - \frac{\theta}{\theta^{*}}\right) + s(Q, \bm{X}^{d}_{-\mathcal{I}}) +  Qs'(Q, \bm{X}^{d}_{-\mathcal{I}}) = MC(Q, \bm{X}^{s}).
\end{align}
As $r(\bm{X}^{d})$ is a function of $X^{d}_i$ by Assumption \ref{assumption:exclusive_shifters}, the equilibrium condition implies that the equilibrium quantity should depend on $X^{d}_i$ under $\theta^{*}$.
Therefore, while the change in $X^{d}_i$ does not change the equilibrium quantity under $\theta$, it changes under $\theta^{*}$, which violates observational equivalence.
Therefore, the conduct parameter and the marginal cost function are always identified.

Recall that Lau's claim includes an edge case where a separable inverse demand function leads to the identification.
The above lemma also includes his edge case when $\mathcal{I} = \emptyset$, that is, all demand shifters work as a demand rotation instrument.
We will see that \eqref{eq:identification_separable_demand} is also the edge case in our result.

\paragraph{Case: The denominators of $C_i$ and $D_i$ are not zero}

Now, suppose that \eqref{eq:identification_condition_separable} does not hold for any $Q$ and $\bm{X}^{d}$ and for all demand shifters.
Then, we take the derivative of $C_i$ and $D_i$ with respect to $X^{d}_j$ for all $j = 1,\ldots, K_d$ and put the derivative being equal to zero.
This process leads to differential equations for the inverse demand function.
Fortunately, these differential equations have analytical solutions, and we can characterize an inverse demand function that satisfies the independence of $C_i$ and $D_i$ of the demand shifter:
\begin{lemma}\label{lemma:nonidentification_inverse_demand}
    The non-identification of the conduct parameter and the marginal cost function implies \eqref{eq:identification_condition_separable} does not hold for all $i  = 1, \ldots, K_d$, and $C_i(Q, \bm{X}^{d})$ and $D_i(Q, \bm{X}^{d})$ are independent of $\bm{X}^{d}$ for all $i$.    
    The independence implies that the inverse demand function is given as
    \begin{align}
        P(Q, \bm{X}^{d}) = Q^{\alpha}r(\bm{X}^{d}) + s(Q)
    \end{align}
    where $\alpha \ne -\frac{1}{\theta}$, and $r: \mathbb{R}^{K_d} \rightarrow \mathbb{R}$ and $s: \mathbb{R}_{+}\rightarrow \mathbb{R}$ are at least twice-continuously differentiable.
\end{lemma}
See Appendix \ref{appendix:proof} for the proof.
Technically, the twice-continuously differentiability of $r$ and $s$ is required to meet Assumption \ref{assumption:three_times_differentiable}.
As we have discussed, this inverse demand function does not allow for any demand rotation instrument.
Recall that \citet{bresnahanOligopoly1982} provides the idea of using a demand rotation instrument to identify the conduct parameter.
Then, \citet{matsumuraResolving2023} formalize his idea by deriving a sufficient condition that guarantees the inverse demand function includes demand rotation instruments for the identification in Bresnahan's setting.
The contraposition of their sufficient condition implies that the non-identification holds only if the inverse demand function does not include any demand rotation instrument.
Therefore, our characterization is the generalization of their result in general settings.

Another difference from Lau's claim is that the conduct parameter can be identified even when the demand shifter is a scalar as long as the inverse demand function includes a demand rotation instrument.
For example, consider an inverse demand function with a scalar demand rotation instrument:
\begin{align}
    P(Q, X^{d}_1) = \alpha_0 + \alpha_1 Q + \alpha_2X^{d}_1 + \alpha_3QX^{d}_1.
\end{align}
In this case, $C_1$ and $D_1$ are given as
\begin{align}
    C_1(Q, X^{d}_1) = \frac{\alpha_2 + \theta\alpha_3Q}{\alpha_2 + \theta^{*}\alpha_3Q}\quad \text{ and }\quad  D_1(Q, X^{d}_1) =  (\theta^{*} - \theta)(-\alpha_1 + \alpha_3X^{d}_1).
\end{align}
This implies that the transformation \eqref{eq:mc_transformation_quantity} is not valid for any marginal cost function because $D_1(Q, X^{d}_1)$ depends on $X^{d}_1$.
Therefore, any $MC^{*}$ that leads to observational equivalence with $MC$ should depend on the demand shifter, under which the exclusion restriction is violated and hence identification is possible.

\begin{remark}
    The differentiability assumptions are imposed on the inverse demand function through Assumption \ref{assumption:three_times_differentiable}. The functions $r$ and $s$ appear as components of the solution to a differential equation implied by the non-identification condition. Their differentiability is therefore inherited from that of $P$ and is not imposed as an independent assumption. Cases in which $r$ or $s$ fail to be sufficiently differentiable are outside the scope of the theorem.  
\end{remark}

\subsection{Sufficient Condition for Non-identification}\label{subsec:sufficient_condition_nonidentification}
We turn to check the sufficiency of the inverse demand function characterized in Lemma \ref{lemma:nonidentification_inverse_demand}.
To see whether the inverse demand function \eqref{eq:nonidentification_inverse_demand} can lead to non-identification, we check whether there exists a transformation between $MC$ and $MC^{*}$ and the marginal revenue under $\theta$ and under $\theta^{*}$ such that any equilibrium in $\mathcal{E}$ can also be an equilibrium in $\mathcal{E}^{*}$ by using the mapping.
Formally, suppose that there exists a transformation $T$ such that
\begin{align}
    MC^{*}(Q, \bm{X}^{s}) = T\left(MC(Q, \bm{X}^{s})\right).
\end{align}
and
\begin{align}
    P(Q, \bm{X}^{d}) + \theta^{*} Q \frac{\partial P}{\partial Q}(Q, \bm{X}^{d}) =T\left(P(Q, \bm{X}^{d}) + \theta Q \frac{\partial P}{\partial Q}(Q, \bm{X}^{d})\right).
\end{align}
Then, suppose also that an equilibrium quantity $Q^e$ satisfies the equilibrium condition under $\mathcal{E}$.
When we have that for $\mathcal{E}^{*}$, 
\begin{align}
    &P(Q^e, \bm{X}^{d}) + \theta^{*} Q^e \frac{\partial P}{\partial Q}(Q^e, \bm{X}^{d})- MC^{*}(Q^e, \bm{X}^{s})\\
    =\ &T\left(P(Q^{e}, \bm{X}^{d}) + \theta Q^{e} \frac{\partial P}{\partial Q}(Q^{e}, \bm{X}^{d})\right) - T\left(MC(Q^{e}, \bm{X}^{s})\right) =0,
\end{align}
then $Q^{e}$ also satisfies the equilibrium condition under $\mathcal{E}^{*}$, and hence the non-identification holds.
In fact, under the inverse demand function \eqref{eq:nonidentification_inverse_demand}, we can construct such transformation $T$ by using \eqref{eq:mc_transformation_quantity}.
Therefore, we can have the following sufficient condition for the non-identification:
\begin{lemma}\label{lemma:sufficient_nonidentification}
    Given the inverse demand function \eqref{eq:nonidentification_inverse_demand} where $\alpha \ne -\frac{1}{\theta}$, given a conduct parameter and a marginal cost function, we can construct another pair of a conduct parameter and a marginal cost function that leads to the same equilibrium point for any value of demand shifters and cost shifters.
    Therefore, non-identification of the conduct parameter and the marginal cost function holds.
\end{lemma}
The proof is given in Appendix \ref{appendix:proof}.
While Lau shows the separable inverse demand function is also a sufficient condition for non-identification, his sufficiency proof is incomplete because it can be verified that his proof implicitly assumes that there exists a transformation $T$ under a separable inverse demand function \eqref{eq:separable_demand}.
However, the separability itself does not guarantee the existence of the transformation.
In contrast, Lemma \ref{lemma:sufficient_nonidentification} shows that the inverse demand function without any demand rotation instrument can lead to a transformation that leads to non-identification.

\subsection{Discussion}\label{subsec:discussion}
\paragraph{Rethinking the role of demand rotation instruments:}
Lemma \ref{lemma:nonidentification_inverse_demand} emphasizes the role of demand rotation instruments.
Intuitively, when a change in a demand shifter keeps the equilibrium quantity the same, we have two equilibrium conditions:
\begin{align}
    P(Q, \bm{X}^{d}) + \theta Q \frac{\partial P}{\partial Q}(Q, \bm{X}^{d}) & = MC(Q, \bm{X}^{s})\\
    P(Q, \tilde{\bm{X}}^{d}) + \theta Q \frac{\partial P}{\partial Q}(Q, \tilde{\bm{X}}^{d}) & = MC(Q, \bm{X}^{s})
\end{align}
Here, we have two equations and two unknowns $(\theta, MC(Q, \bm{X}^{s}))$, and hence we can solve the system of equation with respect to the conduct parameter and the marginal cost function.
Therefore, the existence of a demand shifter that keep the equilibrium quantity the same is confirmed as the sufficient condition for the identification in more general setting.

To build intuition for how the instrument relates to the proof, consider an economy with a linear inverse demand with a demand rotation instrument and a linear marginal cost.
\citet{bresnahanOligopoly1982} considers this environment.
Then, assume that the true conduct is the perfect competition, that is, $\theta = 0$, and the alternative conduct is the monopoly, that is, $\theta^{*} = 1$.
Figure \ref{fig:identification_example} illustrates this environment.
Note that the equilibrium quantity under perfect competition is determined by the intersection of the inverse demand function and the marginal cost function $MC$.
The equilibrium quantity under the monopoly is determined by the intersection of the marginal revenue $MR^{*}$ and the marginal cost function $MC^{*}$.

\begin{figure}[p!]
    \begin{center}
        \begin{subfigure}[b]{0.45\textwidth}
            \centering
            \begin{tikzpicture}[scale = 0.7]
                \draw[->] (-0.5,0) -- (8,0) node[right] {$Q$}; 
                \draw[->] (0,-0.5) -- (0,7.5) node[above] {$P$}; 

                \draw[thick] (0,5) -- (7.5,2.5) node[below right] {$D$};
                \draw[thick] (0,5) -- (5,0) node[below] {$MR^{*}$};

                \draw[thick] (0,1) -- (4.5,5.5) node[above right] {$MC$};

                \draw[thick] (0,1) -- (7.5,3.5) node[above right] {$MC^{*}$};

                \node[circle, fill, inner sep=1.5pt] (E1) at (3,4) {};
                \node[above right] at (E1) {$\quad E$};
                \draw[dashed] (3,0) -- (3,4);
                \draw[dashed] (0,4) -- (3,4);
            \end{tikzpicture}
            \caption{Step 1: Observable equivalence holds.}
            \label{fig:identification_example_step_1}
        \end{subfigure}
        \hfill
        \begin{subfigure}[b]{0.45\textwidth}
            \centering
            \begin{tikzpicture}[scale = 0.7]
                \draw[->] (-0.5,0) -- (8,0) node[right] {$Q$}; 
                \draw[->] (0,-0.5) -- (0,7.5) node[above] {$P$}; 

                \draw[dashed] (0,5) -- (7.5,2.5) ;
                \draw[dashed] (0,5) -- (5,0);

                \draw[thick] (0,7) -- (7,0) node[above right] {$D$};
                \draw[thick] (0,7) -- (3.5,0) node[below] {$MR^{*}$};

                \draw[-<, very thick, red] (0.5,5) to[out=270,in=90] (0.5,5.5);

                \draw[thick] (0,1) -- (4.5,5.5) node[above right] {$MC$};

                \draw[thick] (0,1) -- (7.5,3.5) node[above right] {$MC^{*}$};

                \node[circle, fill, inner sep=1.5pt] (E1) at (3,4) {};
                \node[above right] at (E1) {$\quad E$};
                \draw[dashed] (3,0) -- (3,4);

                \node[circle, fill, inner sep=1.5pt] (E2) at (18/7,7 - 18/7) {};
                \node[above right] at (E2) {$E^{*}$};
                \draw[dashed] (18/7,0) -- (18/7,7 - 18/7);
            \end{tikzpicture}
            \caption{Step 2: Demand rotation changes $D$ and $MR^{*}$}
            \label{fig:identification_example_step_2}
        \end{subfigure}
        \begin{subfigure}[b]{0.45\textwidth}
            \centering
            \begin{tikzpicture}[scale = 0.7]
                \draw[->] (-0.5,0) -- (8,0) node[right] {$Q$}; 
                \draw[->] (0,-0.5) -- (0,7.5) node[above] {$P$}; 

                \draw[dashed] (0,5) -- (7.5,2.5) ;
                \draw[dashed] (0,5) -- (5,0);

                \draw[thick] (0,7) -- (7,0) node[above right] {$D$};
                \draw[thick] (0,7) -- (3.5,0) node[below] {$MR^{*}$};

                \draw[thick] (0,1) -- (4.5,5.5) node[above right] {$MC$};

                \draw[dashed] (0,1) -- (7.5,3.5);
                \draw[->, very thick, red] (6.5,3.5) -- (6.5,2.55);

                \draw[thick] (0,0.0) -- (7.5,2.5) node[below right] {$MC^{*}$};

                \node[circle, fill, inner sep=1.5pt] (E1) at (3,4) {};
                \node[above right] at (E1) {$\quad E = E^{*}$};
                \draw[dashed] (3,0) -- (3,4);
            \end{tikzpicture}
            \caption{Step 3.1: The intercept of $MC^{*}$ changes along with the demand rotation to keep $E = E^{*}$.}
            \label{fig:identification_example_step_3_1}
        \end{subfigure}
        \hfill
        \begin{subfigure}[b]{0.45\textwidth}
            \centering
            \begin{tikzpicture}[scale = 0.7]
                \draw[->] (-0.5,0) -- (8,0) node[right] {$Q$}; 
                \draw[->] (0,-0.5) -- (0,7.5) node[above] {$P$}; 

                \draw[dashed] (0,5) -- (7.5,2.5) ;
                \draw[dashed] (0,5) -- (5,0);

                \draw[thick] (0,7) -- (7,0) node[above right] {$D$};
                \draw[thick] (0,7) -- (3.5,0) node[below] {$MR^{*}$};

                \draw[thick] (0,1) -- (4.5,5.5) node[above right] {$MC$};

                \draw[dashed] (0,1) -- (7.5,3.5);
                \draw[->, very thick, red] (6.5,3) -- (6.5,1.5);

                \draw[thick] (0,1) -- (7,0.9) node[below left] {$MC^{*}$};

                \node[circle, fill, inner sep=1.5pt] (E1) at (3,4) {};
                \node[above right] at (E1) {$\quad E = E^{*}$};
                \draw[dashed] (3,0) -- (3,4);

            \end{tikzpicture}
            \caption{Step 3.2: The slope of $MC^{*}$ changes along with the demand rotation to keep $E = E^{*}$.}
            \label{fig:identification_example_step_3_2}
        \end{subfigure}
    \end{center}
    \caption{Intuition of the role of the demand rotation instrument and identification}
    \label{fig:identification_example}
    \vspace{2mm}
    \footnotesize
    Note: The figures illustrate the intuition of how the demand rotation instrument works to identify the conduct parameter.
    $MC$ is the true marginal cost function, and $MC^{*}$ is the marginal cost function that rationalizes the monopoly conduct.
    Step 1 illustrates that the monopoly and the perfect competition are observationally equivalent.
    In step 2, the demand rotation changes the intercept and slope of the demand function without changing the equilibrium point under the perfect competition, but changes the equilibrium point under the monopoly.
    Step 3 illustrates that to keep the equilibrium point under the monopoly, $MC^{*}$ should change along with the demand rotation, which is impossible because the marginal cost function is independent of the demand shifter.
\end{figure}

In Figure \ref{fig:identification_example_step_1}, the perfect competition and the monopoly lead to the same equilibrium $E$.
Then, consider the change in a demand rotation instrument.
In Figure \ref{fig:identification_example_step_2}, the demand rotation instrument changes the intercept and slope of the demand function without changing the equilibrium point under the true model.
Under the monopoly, the new equilibrium point is $E^{*}$, which is different from $E$, and hence the observational equivalence is violated.
To obtain the same equilibrium after the change in the demand rotation instrument ($E = E^{*}$), $MC^{*}$ should shift as in Figures \ref{fig:identification_example_step_3_1} and \ref{fig:identification_example_step_3_2}.
Note that we changed only the demand shifter, and hence the cost shifter is unchanged.
Thus, for the observational equivalence to hold, the shift in $MC^{*}$ should be derived from the change in the demand rotation instrument, which implies that $MC^{*}$ is a function of the demand shifter.
However, this is impossible because it violates Assumption \ref{assumption:exclusive_shifters}, that is, the demand shifter should not affect the marginal cost function.
The transformation in Lemma \ref{lemma:mc_transformation} clearly shows the possibility of the dependence of the marginal cost function on demand shifters, and hence we need to shut down the effect of demand shifters, which leads to the inverse demand function \eqref{eq:nonidentification_inverse_demand}.

\paragraph{The technical problem in Lau's proof:}
To show Claim \ref{theorem_lau}, Lau follows the same logic as in Lemma \ref{lemma:nonidentification_characterization}, but he obtains a slightly different equation from \eqref{eq:nonidentification_marginal_cost}:
\begin{align}
    \frac{\partial MC^{*}}{\partial X^{s}_j}(Q, \bm{X}^{s}) = \mu(Q, \bm{X}^{s}) \frac{\partial MC}{\partial X^{s}_j}(Q, \bm{X}^{s}),\quad \forall Q, \bm{X}^{s}, \label{eq:nonidentification_marginal_cost_lau}
\end{align}
where $\mu(Q, \bm{X}^{s})$ depends only on $Q$ and $\bm{X}^{s}$.
This corresponds to Equation (15) in \citet{lauIdentifying1982}.
Then he concludes that there is a transformation $T$ such that 
\begin{align}
    MC^{*}(Q^e, \bm{X}^{s}) = T(MC(Q^e, \bm{X}^{s}), Q^e). \label{eq:mc_transformation_lau}
\end{align}
The existence of the transformation $T$ is shown by integrating \eqref{eq:nonidentification_marginal_cost_lau} with respect to the cost shifter when it is a scalar, and by applying Lemma \ref{lemma_1_GU} from \citet{goldmanNote1964} when the cost shifter is a vector.
Note also that in our approach, both \eqref{eq:nonidentification_demand} and \eqref{eq:nonidentification_marginal_cost} are necessary to derive the transformation between marginal cost functions. 
In contrast, while Lau derives both equations \eqref{eq:nonidentification_demand} and \eqref{eq:nonidentification_marginal_cost} in his proof, he relies solely on \eqref{eq:nonidentification_marginal_cost} to show the non-identification.

While Lau does not explicitly define $\mu$ in his formulation, it can be verified that his $\mu$ corresponds to our $\lambda$.
However, he neglects a point that the function $\mu$ depends on the demand shifter $\bm{X}^{d}$, and this dependence is not addressed in his analysis.
As long as the effect of the demand shifter cannot be eliminated from $\mu$, the transformation $T$ between $MC$ and $MC^*$ could depend on $\bm{X}^{d}$, which violates Assumption \ref{assumption:exclusive_shifters}.
Additionally, separability does not resolve the dependence of $\mu$ on the demand shifter.
Unfortunately, Lau does not provide a justification for why he can assume that the effect of the demand shifter can be removed from $\mu$, and hence it leaves a critical gap in the argument.
Additionally, Lau only assumes that the inverse demand functions are twice continuously differentiable, which allows him to apply Lemma \ref{lemma_1_GU} to justify the existence of the transformation $T$. 
However, to eliminate the effect of $\bm{X}^{d}$ from $\mu$, it is necessary to differentiate $\mu$ with respect to $\bm{X}^{d}$ and require that the derivative is equal to zero as in our analysis.
This step, in turn, requires that the inverse demand function be at least three-times continuously differentiable as in our proof.
Therefore, twice-continuous differentiability is not enough to characterize the non-identification.

\section{Other Discussions}\label{sec:other_discussions}
\paragraph{Relationship with Firm Conduct Test}

In recent years, several papers have developed tests for firm conduct \citep*{backusCommon2021,duarteTesting2024} that compare two different models and statistically determine which firm conduct is close to the true firm conduct.
\citet{dearingLearning2024} investigate the mechanism how instrument variables can distinguish the true firm conduct by focusing on pass-through.
\citet{magnolfiComparison2022} clarify when firm conduct test outperforms firm conduct estimation.
While these papers consider differentiated-product markets, the logic can be applied to homogeneous-product markets, and our result emphasizes that demand rotation instruments are essential in firm conduct test.

The results so far hold whenever we adopt the conduct parameter approach.
However, the conduct parameter approach has been criticized on several fronts.
In this section, we discuss its micro-foundations and its empirical accuracy.

\paragraph{Micro-foundations of Conduct Parameter Approach}
The conduct parameter approach is based on the conjectural variation model.
In the conjectural variation model, each firm has a conjecture about how the competitors will react to the firm's action.
A major problem of the conjectural variation model is that the conjecture and the reaction function do not coincide at equilibrium.
\citet{bresnahanDuopoly1981,bresnahanExistence1983}  propose a Consistent Conjecture Equilibrium (CCE) that requires the conjecture and the reaction function to coincide at equilibrium.
However, the existence and uniqueness of the CCE are not guaranteed in general \citep{klempererConsistent1988,robsonExistence1983}.

Several alternative micro-foundations have been proposed.
For example, \citet{escrihuela-villarNote2015} shows that the corporation coefficient approach is equivalent to the conjectural variation model in some specific settings.
In this model, each firm maximizes its profit plus the weighted sum of its competitors' profits, where the weight is called the corporation coefficient.
\citet{menezesStrategic2020} investigates the supply function approach in which firms choose not a quantity but a supply schedule based on price.
\citet{menezesCompetition2023} show the relationship between the conduct parameter approach and the supply function approach in a model with linear demand and linear marginal cost, and \citet{menezesStrategic2020} briefly discuss how to estimate competitiveness given information on marginal cost.


\paragraph{Accuracy of the Conduct Parameter Approach}
While this paper focuses only on the identification problem, \citet{cortsConduct1999} argues that the conduct parameter approach can be inaccurate when the data-generating process lies outside the conduct parameter model.
For example, while a repeated game model can sustain the equilibrium quantity under joint-profit maximization, the conduct parameter estimate could be biased away from $\theta = 1$.
As \citet{cortsConduct1999} and \citet{magnolfiComparison2022} mention, this criticism can be mitigated by assuming that the data-generating process is a static competition derived from the equilibrium condition in \eqref{eq:foc}.

\section{Conclusion}\label{sec:conclusion}

This paper revisited the identification of the conduct parameter in homogeneous product markets.
In the literature, Lau considers the identification problem in a fairly generalized setting.
While \citet{lauIdentifying1982} characterizes the non-identification condition, we point out some problems in Lau's paper and provide a novel characterization of the non-identification condition.
Based on the new characterization, we find that the non-identification is equivalent to the absence of demand shifters that work as demand rotation instruments proposed in \citet{bresnahanOligopoly1982}.
We then consider the identification of all primitives in the model and show that the inverse demand function that leads to the non-identification of the conduct parameter and the marginal cost function cannot be identified, which immediately implies that the conduct parameter and the marginal cost function are also not identified.
Therefore, while a demand rotation instrument is key for identifying the conduct parameter, identifying the entire model requires an inverse demand function with a demand rotation instrument that is itself identifiable.

\bibliographystyle{aer}
\bibliography{conduct_parameter_counter_example}

\newpage
\appendix
\appendixsection
\numberwithin{theorem}{section}
\numberwithin{lemma}{section}
\numberwithin{proposition}{section}
\numberwithin{corollary}{section}
\numberwithin{definition}{section}
\numberwithin{example}{section}
\numberwithin{assumption}{section}
\numberwithin{claim}{section}

\begin{center}
\huge\textbf{Appendix}
\end{center}
\vspace{1mm}

\section{Omitted Proofs}\label{appendix:proof}
In this section, we provide the proofs of the lemmas and theorems in the main text.
In what follows, we suppress the arguments of functions when no confusion arises.

\subsection{Proof of Lemma \ref{lemma:nonidentification_characterization}}
Assume that given $\bm{X}^{d}$ and $\bm{X}^{s}$, we can solve the equilibrium condition for the equilibrium quantity $Q^e$.
The non-identification implies that the reduced form quantity $Q^{e} = h_q(\bm{X}^{d}, \bm{X}^{s}) = h^{*}_q(\bm{X}^{d}, \bm{X}^{s}) $ satisfies 
\begin{align}
    P(Q^e, \bm{X}^{d}) + \theta Q^e\frac{\partial P}{\partial Q}(Q^e, \bm{X}^{d}) & = MC(Q^e, \bm{X}^{s}).  \label{eq:foc_theta}
\end{align}
and
\begin{align}
    P(Q^e, \bm{X}^{d}) + \theta^{*} Q^e\frac{\partial P}{\partial Q}(Q^e, \bm{X}^{d}) & = MC^{*}(Q^e, \bm{X}^{s}), \label{eq:foc_theta_star}
\end{align}
for any $\bm{X}^{d}$ and $\bm{X}^{s}$.  
For notational simplicity, we define
\begin{align}
    F(Q^e, \bm{X}^{d}, \bm{X}^{s}; \theta, MC) \equiv P(Q^e, \bm{X}^{d}) + \theta Q^e\frac{\partial P}{\partial Q}(Q^e, \bm{X}^{d}) - MC(Q^e, \bm{X}^{s}).
\end{align}
Then, by Assumption \ref{assumption:unique_equilibrium}, we can apply the implicit function theorem, and for the reduced-form equation $Q^e = h_q(\bm{X}^{d}, \bm{X}^{s})$, the gradient of the reduced-form function $h_q$ with respect to $\bm{X}^{d}$ and $\bm{X}^{s}$ is given by
\begin{align}
    \nabla h_q(\bm{X}^{d}, \bm{X}^{s}) =  \left( -\dfrac{\frac{\partial F}{\partial X^{m}_{i}}(h_q(\bm{X}^{d}, \bm{X}^{s}), \bm{X}^{d}, \bm{X}^{s})}{\frac{\partial F}{\partial Q}(h_q(\bm{X}^{d}, \bm{X}^{s}), \bm{X}^{d}, \bm{X}^{s})} \right)_{\substack{m = d, s\\ i = 1, \ldots, K_m}}. \label{eq:foc_derivative_demand_supply}
\end{align}
The derivatives of $F$ for each variable are for $i = 1, \ldots, K_d$ and $i = 1, \ldots, K_s$,
\begin{align}
    \frac{\partial F}{\partial X^{d}_i}(Q^e, \bm{X}^{d}, \bm{X}^{s}; \theta, MC) & =  \frac{\partial P}{\partial X^{d}_i}(Q^e, \bm{X}^{d}) + \theta\frac{\partial^2 P}{\partial X^{d}_i\partial Q}(Q^e, \bm{X}^{d})Q^e,\\
    \frac{\partial F}{\partial X^{s}_i}(Q^e, \bm{X}^{d}, \bm{X}^{s}; \theta, MC) & =  -\frac{\partial MC}{\partial X^{s}_{i}}(Q^e, \bm{X}^{s}), \\
    \frac{\partial F}{\partial Q}(Q^e, \bm{X}^{d}, \bm{X}^{s}; \theta, MC) & = (1+\theta)\frac{\partial P}{\partial Q}(Q^e, \bm{X}^{d}) + \theta Q^e\frac{\partial^2 P}{\partial Q^2}(Q^e, \bm{X}^{d}) - \frac{\partial MC}{\partial Q}(Q^e, \bm{X}^{s}).
\end{align}
Thus, the derivative of $h$ with respect to $X^{d}_i$ and $X^{s}_i$ are given as for $i = 1, \ldots, K_d$ and $i = 1, \ldots, K_s$,
\begin{align}
    \frac{\partial h_q}{\partial X^{d}_i}(\bm{X}^{d}, \bm{X}^{s}) = -\frac{\frac{\partial P}{\partial X^{d}_i}(Q^e, \bm{X}^{d}) + \theta Q^e \frac{\partial^2 P}{\partial X^{d}_i\partial Q}(Q^e, \bm{X}^{d}) }{(1+\theta)\frac{\partial P}{\partial Q}(Q^e, \bm{X}^{d}) + \theta  Q^e\frac{\partial^2 P}{\partial Q^2}(Q^e, \bm{X}^{d}) - \frac{\partial MC}{\partial Q}(Q^e, \bm{X}^{s})}, \label{eq:foc_derivative_demand}
\end{align}
and
\begin{align}
    \frac{\partial h_q}{\partial X^{s}_{i}}(\bm{X}^{d}, \bm{X}^{s}) & = \frac{\frac{\partial MC}{\partial X^{s}_{i}}(Q^e, \bm{X}^{s})}{(1+\theta)\frac{\partial P}{\partial Q}(Q^e, \bm{X}^{d}) + \theta  Q^e\frac{\partial^2 P}{\partial Q^2}(Q^e, \bm{X}^{d}) - \frac{\partial MC}{\partial Q}(Q^e, \bm{X}^{s})}. \label{eq:foc_derivative_supply}
\end{align}

Note that the same argument can be applied to the equilibrium condition \eqref{eq:foc_theta_star} of the alternative model.
Recall that the non-identification implies that $Q^e = h_q(\bm{X}^{d}, \bm{X}^{s}) = h_q^{*}(\bm{X}^{d}, \bm{X}^{s})$ for all $\bm{X}^{d}$ and $\bm{X}^{s}$, and hence we must have
\begin{align}
    \nabla h_q^{*}(\bm{X}^{d}, \bm{X}^{s}) = \nabla h_q(\bm{X}^{d}, \bm{X}^{s}) \quad \forall \bm{X}^{d}, \bm{X}^{s}. \label{eq:observable_equivalence_derivative}
\end{align}
From \eqref{eq:foc_derivative_demand_supply}, this implies that for all $Q^e$, $\bm{X}^{d}$, and $\bm{X}^{s}$,
\begin{align}
    \begin{pmatrix}
        \frac{\partial F}{\partial \bm{X}^{d}_{1}}(Q^e, \bm{X}^{d}, \bm{X}^{s}; \theta^{*}, MC^{*})\\
        \vdots \\
        \frac{\partial F}{\partial \bm{X}^{d}_{K_d}}(Q^e, \bm{X}^{d}, \bm{X}^{s}; \theta^{*}, MC^{*})\\
        \frac{\partial F}{\partial X^{s}_{1}}(Q^e, \bm{X}^{d}, \bm{X}^{s}; \theta^{*}, MC^{*})\\
        \vdots \\
        \frac{\partial F}{\partial X^{s}_{K_s}}(Q^e, \bm{X}^{d}, \bm{X}^{s}; \theta^{*}, MC^{*})
    \end{pmatrix}
    = \lambda(Q^e, \bm{X}^{d}, \bm{X}^{s})
    \begin{pmatrix}
        \frac{\partial F}{\partial \bm{X}^{d}_{1}}(Q^e, \bm{X}^{d}, \bm{X}^{s}; \theta, MC)\\
        \vdots \\
        \frac{\partial F}{\partial \bm{X}^{d}_{K_d}}(Q^e, \bm{X}^{d}, \bm{X}^{s}; \theta, MC)\\
        \frac{\partial F}{\partial X^{s}_{1}}(Q^e, \bm{X}^{d}, \bm{X}^{s}; \theta, MC)\\
        \vdots \\
        \frac{\partial F}{\partial X^{s}_{K_s}}(Q^e, \bm{X}^{d}, \bm{X}^{s}; \theta, MC)
    \end{pmatrix},\label{eq:foc_derivative_demand_supply_lambda}
\end{align}
where $\lambda(Q^e, \bm{X}^{d}, \bm{X}^{s})$ is defined as
\begin{align}
    \lambda(Q^e, \bm{X}^{d}, \bm{X}^{s}) \equiv \frac{(1+\theta^{*})\frac{\partial P}{\partial Q}(Q^e, \bm{X}^{d}) + \theta^{*} Q^e\frac{\partial^2 P}{\partial Q^2}(Q^e, \bm{X}^{d}) - \frac{\partial MC^{*}}{\partial Q}(Q^e, \bm{X}^{s})}{(1+\theta)\frac{\partial P}{\partial Q}(Q^e, \bm{X}^{d}) + \theta Q^e\frac{\partial^2 P}{\partial Q^2}(Q^e, \bm{X}^{d}) - \frac{\partial MC}{\partial Q}(Q^e, \bm{X}^{s})}.
\end{align}
Assumption \ref{assumption:unique_equilibrium} guarantees that $\lambda(Q^e, \bm{X}^{d}, \bm{X}^{s})$ is nonzero and finite.
From the first $K_d$ rows in \eqref{eq:foc_derivative_demand_supply_lambda}, \eqref{eq:nonidentification_demand} and for the later $K_s$ rows, \eqref{eq:nonidentification_marginal_cost} hold.

\subsection{Proof of Lemma \ref{lemma:mc_transformation}}

From \eqref{eq:nonidentification_demand}, we have (we suppress the arguments of the functions for brevity)
\begin{align}
    &\left[\frac{\partial P}{\partial X^{d}_i} + \theta^{*} Q^e\frac{\partial^2 P}{\partial X^{d}_i\partial Q}\right]\left[(1+\theta)\frac{\partial P}{\partial Q} + \theta Q^e\frac{\partial^2 P}{\partial Q^2} - \frac{\partial MC}{\partial Q}\right] \\
    & \hspace{4cm}= \left[\frac{\partial P}{\partial X^{d}_i} + \theta Q^e\frac{\partial^2 P}{\partial X^{d}_i\partial Q}\right]\left[(1+\theta^{*})\frac{\partial P}{\partial Q} + \theta^{*} Q^e\frac{\partial^2 P}{\partial Q^2} - \frac{\partial MC^{*}}{\partial Q}
    \right]\\
    & \frac{\frac{\partial P}{\partial X^{d}_i} + \theta^{*} Q\frac{\partial^2 P}{\partial X^{d}_i\partial Q} }{\frac{\partial P}{\partial X^{d}_i} + \theta Q\frac{\partial^2 P}{\partial X^{d}_i\partial Q} }\left[(1+\theta)\frac{\partial P}{\partial Q} + \theta Q^e\frac{\partial^2 P}{\partial Q^2} - \frac{\partial MC}{\partial Q}\right]\\
    &\hspace{4cm} =\left[(1+\theta^{*})\frac{\partial P}{\partial Q} + \theta^{*} Q^e\frac{\partial^2 P}{\partial Q^2} - \frac{\partial MC^{*}}{\partial Q}
    \right]\\
    & \frac{\partial MC^{*}}{\partial Q} = \underbrace{(1+\theta^{*})\frac{\partial P}{\partial Q} + \theta^{*} Q^e\frac{\partial^2 P}{\partial Q^2} - C_i(Q, \bm{X}^{d}) \left[(1+\theta)\frac{\partial P}{\partial Q} + \theta Q^e\frac{\partial^2 P}{\partial Q^2} \right]}_{(\ast)} + C_i(Q, \bm{X}^{d})\frac{\partial MC}{\partial Q}.
\end{align}
The terms in $(\ast)$ can be more simplified as 
\begin{align}
    &(1+ \theta^{*})\frac{\partial P}{\partial Q} + \theta^{*} Q\frac{\partial^2 P}{\partial Q^2} - \frac{\frac{\partial P}{\partial X^{d}_i} + \theta^{*} Q\frac{\partial^2 P}{\partial X^{d}_i\partial Q} }{\frac{\partial P}{\partial X^{d}_i} + \theta Q\frac{\partial^2 P}{\partial X^{d}_i\partial Q} }\left[(1+ \theta) \frac{\partial P}{\partial Q} + \theta Q\frac{\partial^2 P}{\partial Q^2}\right]\\
    &= \frac{1}{\frac{\partial P}{\partial X^{d}_i} + \theta\frac{\partial^2 P}{\partial X^{d}_i\partial Q}Q}\Bigg[\left((1 + \theta^{*}) \frac{\partial P}{\partial Q} + \theta^{*} Q\frac{\partial^2 P}{\partial  Q^2}\right)\left(\frac{\partial P}{\partial X^{d}_i} + \theta Q\frac{\partial^2 P}{\partial X^{d}_i\partial Q}\right)\\
    &\hspace{4cm} - \left( (1 + \theta) \frac{\partial P}{\partial Q} + \theta Q\frac{\partial^2 P}{\partial Q^2}\right)\left(\frac{\partial P}{\partial X^{d}_i} + \theta^{*} Q\frac{\partial^2 P}{\partial X^{d}_i\partial Q}\right)\Bigg]\\
    & =  \frac{\theta^{*} - \theta}{\frac{\partial P}{\partial X^{d}_i} + \theta\frac{\partial^2 P}{\partial X^{d}_i\partial Q}Q}\left[ \frac{\partial P}{\partial Q} \frac{\partial P}{\partial X^{d}_i} + Q\frac{\partial P}{\partial X^{d}_i}\frac{\partial^2 P}{\partial^2 Q} - Q\frac{\partial P}{\partial Q}\frac{\partial^2 P}{\partial X^{d}_i \partial Q} \right],
\end{align}
which corresponds to $D_i(Q, \bm{X}^{d})$ in \eqref{eq:interaction_derivative_demand}, and hence we have
\begin{align}
    \frac{\partial MC^{*}}{\partial Q}(Q, \bm{X}^{s}) = D_i(Q, \bm{X}^{d}) + C_i(Q, \bm{X}^{d})\frac{\partial MC}{\partial Q}(Q, \bm{X}^{s}).
\end{align}

Next, \eqref{eq:nonidentification_demand} and \eqref{eq:nonidentification_marginal_cost} imply that for $i = 1, \ldots, K_d$ and $j = 1, \ldots, K_s$,
\begin{align}
    \lambda(Q^e, \bm{X}^{d}, \bm{X}^{s}) =  \frac{\frac{\partial MC^{*}}{\partial X^{s}_j}(Q^e, \bm{X}^{s})}{\frac{\partial MC}{\partial X^{s}_j}(Q^e, \bm{X}^{s})} =  \frac{\frac{\partial P}{\partial X^{d}_i}(Q^e, \bm{X}^{d}) + \theta^{*} Q^e\frac{\partial^2 P}{\partial X^{d}_i\partial Q}(Q^e, \bm{X}^{d}) }{\frac{\partial P}{\partial X^{d}_i}(Q^e, \bm{X}^{d}) + \theta Q^e\frac{\partial^2 P}{\partial X^{d}_i\partial Q}(Q^e, \bm{X}^{d})} \equiv C_i(Q, \bm{X}^{d}),
\end{align}
Then, we have 
\begin{align}
    \frac{\partial MC^{*}}{\partial X^{s}_j}(Q, \bm{X}^{s}) = C_i(Q, \bm{X}^{d}) \frac{\partial MC}{\partial X^{s}_j}(Q, \bm{X}^{s}).
\end{align}

\subsection{Proof of Lemma \ref{lemma:identification_condition_separable}}

Let $u_i(Q, \bm{X}^{d}) \equiv \frac{\partial P}{\partial X^{d}_i}(Q, \bm{X}^{d})$.
Due to Assumption \ref{assumption:effectiveness_shifters}, we have that $u_i(Q, \bm{X}^{d}) \ne 0$ for all $Q$ and $\bm{X}^{d}$.
Note that when $\theta = 0$, \eqref{eq:identification_condition_separable} implies that $u_i(Q, \bm{X}^{d}) = 0$ for all $i \in \mathcal{I}$, which cannot be held because of Assumption \ref{assumption:effectiveness_shifters}.
Therefore, $\theta \ne 0$ to have \eqref{eq:identification_condition_separable}.
Additionally, \eqref{eq:identification_condition_separable} is reduced to $u_i(0, \bm{X}^{d}) =0$ when $ Q = 0$.
Again, this cannot hold because of Assumption \ref{assumption:effectiveness_shifters}.
Thus, we characterize the inverse demand function where \eqref{eq:identification_condition_separable} holds only on $Q >0$ and $\theta \ne 0$.

Rewrite \eqref{eq:identification_condition_separable} for $i \in \mathcal{I}$ as
\begin{align}
    \frac{\frac{\partial u_i}{\partial Q}(Q, \bm{X}^{d})}{ u_i(Q, \bm{X}^{d}) } = - \frac{1}{\theta Q} \Longrightarrow \frac{\partial }{\partial Q}\log |u_i(Q, \bm{X}^{d})| = -\frac{1}{\theta Q}.
\end{align}
By integrating both sides with respect to $Q$, we have
\begin{align}
    \log |u_i(Q, \bm{X}^{d})| = -\frac{1}{\theta}\log Q + R_i(\bm{X}^{d}),
\end{align}
where $R_i(\bm{X}^{d})$ is a function of $\bm{X}^{d}$.
By taking the exponential of both sides, we have
\begin{align}
    |u_i(Q, \bm{X}^{d})| = Q^{-\frac{1}{\theta}}r_i(\bm{X}^{d}),
\end{align}
where $r_i(\bm{X}^{d})  = \exp(R_i(\bm{X}^{d}))$.
Note that we can remove the absolute value for $Q$ because it can be assumed that $Q\ge 0$.
By removing the absolute value operator for $u_i$, we have two solutions
\begin{align}
    u_i(Q, \bm{X}^{d}) = \pm Q^{-\frac{1}{\theta}}r_i(\bm{X}^{d}). 
\end{align}
Because $r_i$ is an arbitrary function, we can unify these solutions and can simply put the solution as
\begin{align}
    u_i(Q, \bm{X}^{d}) = Q^{-\frac{1}{\theta}}r_i(\bm{X}^{d}). 
\end{align}
This implies that the derivative of $P$ with respect to $X^{d}_i$ is a separable function of $Q$ and $\bm{X}^{d}$.
Hence, it is natural to think that $r_i(\bm{X}^{d})$ is a derivative of a function of $\bm{X}^{d}$ with respect to $X^{d}_i$.
In other words, there is a function $r(\bm{X}^{d})$ of $\bm{X}^{d}$ such that $r_i(\bm{X}^{d}) = \frac{\partial r(\bm{X}^{d})}{\partial X^{d}_i}$ holds.
Therefore, we have for $i \in \mathcal{I}$,
\begin{align}
    u_i(Q, \bm{X}^{d}) = \frac{\partial r(\bm{X}^{d})}{\partial X^{d}_i} Q^{-\frac{1}{\theta}}.
\end{align}

By integrating both sides with respect to $X^{d}_i$, we have
\begin{align}
    P(Q, \bm{X}^{d}) = Q^{-\frac{1}{\theta}}r(\bm{X}^{d}) + s_i(Q, \bm{X}^{d}_{-i}), \quad i \in \mathcal{I}, \label{eq:identification_separable_demand_i}
\end{align}
where $s_i(Q, \bm{X}^{d}_{-i})$ is a function of $Q$ and $\bm{X}^{d}_{-i}$ where $\bm{X}^{d}_{-i}$ is the vector of $\bm{X}^{d}$ excluding $X^{d}_i$.
To meet Assumption \ref{assumption:three_times_differentiable}, we must have that $r(\bm{X}^{d})$ and $s_i(Q, \bm{X}^{d}_{-i})$ are at least twice-continuously differentiable.

When $\bm{X}^{d}$ is a scalar or when $\bm{X}^{d}$ is a vector but $|\mathcal{I}| \ge K_d-1$, there is no argument in $\bm{X}^{d}_{-i}$, and hence we can remove the index of $i$ from the function $s_i$.
Therefore, $s(Q, \bm{X}^{d}_{-i}) = s(Q)$ holds, and hence we have \eqref{eq:identification_separable_demand}.
When $\bm{X}^{d}$ is a vector and $K_d -2 \ge|\mathcal{I}|\ge 2$, we can remove the index $i$ from $s_i$ in the following way.
Pick up any $i$ and $j$ in $\mathcal{I}$ such that $i \ne j$, and then the derivative of \eqref{eq:identification_separable_demand_i} with respect to $X^{d}_j$ is given by
\begin{align}
    \frac{\partial P}{\partial X^{d}_j}(Q, \bm{X}^{d}) = \frac{\partial r(\bm{X}^{d})}{\partial X^{d}_j} Q^{-\frac{1}{\theta}} + \frac{\partial s_i(Q, \bm{X}^{d}_{-i})}{\partial X^{d}_j},
\end{align}
and
\begin{align}
    \frac{\partial P}{\partial X^{d}_j}(Q, \bm{X}^{d}) = \frac{\partial r(\bm{X}^{d})}{\partial X^{d}_j} Q^{-\frac{1}{\theta}}.
\end{align}
Since $s_{j}$ is a function of the demand shifters without $X_{j}$, we do not have the term relating to $s_{j}$ in the second equation.
Therefore, by comparing the above two equations, we can conclude that the second term in the first equation vanishes:
\begin{align}
    \frac{\partial s_i(Q, \bm{X}^{d}_{-i})}{\partial X^{d}_j} = 0,
\end{align} 
which implies that $s_i(Q, \bm{X}^{d}_{-i})$ is independent of $X^{d}_j$.
By applying the same argument to all $i \in \mathcal{I}$, we can show that $s_i(Q, \bm{X}^{d}_{-i})$ is independent of the demand shifters whose indices are in $\mathcal{I}$.
That is, we can reduce $s_i(Q, \bm{X}^{d}_{-i})$ to $s_i(Q, \bm{X}^{d}_{-\mathcal{I}})$ where $\bm{X}^{d}_{-\mathcal{I}}$ is the vector of $\bm{X}^{d}$ excluding the demand shifters whose indices are in $\mathcal{I}$.
Then, by comparing \eqref{eq:identification_separable_demand_i} for all $i \in \mathcal{I}$, we have
\begin{align}
    s_i(Q, \bm{X}^{d}_{-\mathcal{I}}) = s_j(Q, \bm{X}^{d}_{-\mathcal{I}}),
\end{align}
for all $i,j \in \mathcal{I}$.
This leads to the symmetry of $s_i$ for all $i \in \mathcal{I}$,
which implies that $s(Q, \bm{X}^{d}_{-\mathcal{I}}) = s_i(Q, \bm{X}^{d}_{-\mathcal{I}})$ for all $i \in \mathcal{I}$.
Therefore, we can write the inverse demand function as
\begin{align}
    P(Q, \bm{X}^{d}) = Q^{-\frac{1}{\theta}}r(\bm{X}^{d}) + s(Q, \bm{X}^{d}_{-\mathcal{I}}).
\end{align}
When $|\mathcal{I}| = 1$, we have $\bm{X}^{d}_{-\mathcal{I}} = \bm{X}^{d}_{-i}$, and hence we can keep the index for $s_i$, but it is also fine to write it as $s(Q, \bm{X}^{d}_{-i}) = s_i(Q, \bm{X}^{d}_{-i})$.

Note that when $Q=0$, we have that
\begin{align}
    u_i(0, \bm{X}^{d}) = 0^{-\frac{1}{\theta}}r_i(\bm{X}^{d}) = \infty,
\end{align}
because $-\frac{1}{\theta} <0$.
This is consistent with the argument that \eqref{eq:identification_condition_separable} does not hold at $Q = 0$.
Thus, we have characterized the inverse demand function where \eqref{eq:identification_condition_separable} holds.

\subsection{Proof of Lemma \ref{lemma:nonidentification_inverse_demand}}
Because $C_i(Q, \bm{X}^{d})$ and $D_i(Q, \bm{X}^{d})$ should be independent of $\bm{X}^{d}$ simultaneously, we can first characterize the class of inverse demand functions that make $C_i(Q, \bm{X}^{d})$ independent of $\bm{X}^{d}$.
Then, we characterize the class of inverse demand functions that make $D_i(Q, \bm{X}^{d})$ independent of $\bm{X}^{d}$ by substituting the derived inverse demand function into $D_i(Q, \bm{X}^{d})$.

\subsubsection*{Step 1: Necessity for $C_i$ is independent of $\bm{X}^{d}$}
When $C_i(Q, \bm{X}^{d})$ is independent of $\bm{X}^{d}$, the derivative of $C_i(Q, \bm{X}^{d})$ with respect to $\bm{X}^{d}$ is zero.
Let $u_i(Q, \bm{X}^{d}) \equiv \frac{\partial P}{\partial X^{d}_i}(Q, \bm{X}^{d})$.
Due to Assumption \ref{assumption:effectiveness_shifters}, we have that $u_i(Q, \bm{X}^{d}) \ne 0$ for all $Q$ and $\bm{X}^{d}$.
Then, the derivative of $C_i$ with respect to $X^{d}_j$ for $j = 1, \ldots, K_d$ is given by
\begin{align}
    \frac{\partial C_i}{\partial X^{d}_j}(Q, \bm{X}^{d}) & = \frac{(\theta^{*} - \theta)Q }{\left(u_i + \theta Q \frac{\partial u_i}{\partial Q}\right)^2}\left[\frac{\partial u_i}{\partial X^{d}_j\partial Q} u_i - \frac{\partial u_i}{\partial Q} \frac{\partial u_i}{\partial X^{d}_j}\right].
\end{align}

Note that because \eqref{eq:identification_condition_separable} does not hold, the denominator is not zero.
As $\theta \ne \theta^{*}$, the derivative becomes zero when (1) $Q = 0$ or (2) the term in the bracket is zero.
In the first case, for arbitrary inverse demand function, $C_i$ is independent of $\bm{X}^{d}$.
Therefore, we focus on the second case for $Q >0$.
Note that the bracket term is a partial differential equation:
\begin{align}
    \frac{\partial u_i}{\partial X^{d}_j\partial Q} u_i - \frac{\partial u_i}{\partial Q} \frac{\partial u_i}{\partial X^{d}_j} = 0. \label{eq:identification_condition_separable_step1}
\end{align}
Now, instead of using \eqref{eq:identification_condition_separable_step1} directly, we use the following relationship:
\begin{align}
    \frac{\partial^2 }{\partial X^{d}_j \partial Q}\log |u_i(Q, \bm{X}^{d})| & = \frac{\partial }{\partial Q}\left(\frac{1}{u_i}\frac{\partial u_i}{\partial X^{d}_j}\right) = \frac{1}{u_i^2}\left(u_i\frac{\partial^2 u_i}{\partial X^{d}_j \partial Q} - \frac{\partial u_i}{\partial X^{d}_j}\frac{\partial u_i}{\partial Q}\right).
\end{align}
Note that the assumption $u_i \ne 0$ implies that the inside of the log function is not zero and the denominator in the last equality is not zero.
Therefore, to check the independence of $C_i$ with respect to $\bm{X}^{d}$, it is sufficient to solve the following partial differential equation such that 
\begin{align}
    \frac{\partial^2 }{\partial X^{d}_j \partial Q}\log |u_i(Q,\bm{X}^{d})| = 0, \quad i,j = 1, \ldots, K_d.
\end{align}
This implies that $\frac{\partial }{\partial Q}\log |u_i(Q, \bm{X}^{d})|$ is independent of any element in $\bm{X}^{d}$, and hence we have a function $G(Q)$ such that
\begin{align}
    \frac{\partial }{\partial Q}\log |u_i(Q, \bm{X}^{d})| = G(Q).
\end{align}

By integrating both sides with respect to $Q$, we have for $i = 1, \ldots, K_d$,
\begin{align}
    \log |u_i(Q, \bm{X}^{d})| = \int G(Q) dQ + R_i(\bm{X}^{d}),
\end{align}
where the last term is a function that is an analogue of the constant of integration.
By taking exponential on both sides, we have for $i = 1, \ldots, K_d$,
\begin{align}
    & |u_i(Q, \bm{X}^{d})| = g(Q)r_i(\bm{X}^{d})
\end{align}
where $r_i(\bm{X}^{d}) \equiv \exp(R_i(\bm{X}^{d}))$ and $g(Q) = \exp\left(\int G(Q) dQ\right)$.
This has two solutions, but again, because $r_i$ is an arbitrary function, we can unify these solutions and put the solution as
\begin{align}
    u_i(Q, \bm{X}^{d}) = g(Q)r_i(\bm{X}^{d}).
\end{align}

Because $u_i$ is a separable function of $Q$ and $\bm{X}^{d}$, it is natural to think that $r_i(\bm{X}^{d})$ is a derivative of a function of $\bm{X}^{d}$ with respect to $X^{d}_i$.
In other words, there is a function $r(\bm{X}^{d})$ of $\bm{X}^{d}$ such that $r_i(\bm{X}^{d}) = \frac{\partial r(\bm{X}^{d})}{\partial X^{d}_i}$ holds.
Therefore, we have
\begin{align}
    u_i(Q, \bm{X}^{d}) = \frac{\partial r(\bm{X}^{d})}{\partial X^{d}_i} g(Q).
\end{align}
Integrating both sides with respect to $X^{d}_i$, we have
\begin{align}
   P(Q, \bm{X}^{d}) = g(Q) r(\bm{X}^{d}) + s_i(Q, \bm{X}^{d}_{-i}), \quad i = 1, \ldots, K_d, \label{eq:inverse_demand_separable_step1}
\end{align}
where $s_i(Q, \bm{X}^{d}_{-i})$ is an arbitrary function of $Q$ and $\bm{X}^{d}_{-i}$ where $\bm{X}^{d}_{-i}$ is the vector of $\bm{X}^{d}$ excluding $X^{d}_i$.
To meet Assumption \ref{assumption:three_times_differentiable}, we must have that $R(\bm{X}^{d})$ and $s_i(Q, \bm{X}^{d}_{-i})$ are at least twice-continuously differentiable and $g(Q)$ is at least continuously differentiable.

We can remove the index of $s_i$ in the following way.
Pick up any $i$, and then the derivative of \eqref{eq:inverse_demand_separable_step1} with respect to $X^{d}_i$ is given by
\begin{align}
    \frac{\partial P}{\partial X^{d}_i}(Q, \bm{X}^{d}) = \frac{\partial r(\bm{X}^{d})}{\partial X^{d}_i} g(Q) + \frac{\partial s_i(Q, \bm{X}^{d}_{-i})}{\partial X^{d}_i} = \frac{\partial r(\bm{X}^{d})}{\partial X^{d}_i} g(Q),
\end{align}
and for any other $j \ne i$, the same derivative is given by
\begin{align}
    \frac{\partial P}{\partial X^{d}_i}(Q, \bm{X}^{d}) = \frac{\partial r(\bm{X}^{d})}{\partial X^{d}_i} g(Q) + \frac{\partial s_j(Q, \bm{X}^{d}_{-j})}{\partial X^{d}_i}.
\end{align}
These imply that
\begin{align}
    \frac{\partial s_j(Q, \bm{X}^{d}_{-j})}{\partial X^{d}_i} = 0 \quad \text{for all } j \ne i.
\end{align}
Therefore, $s_j(Q, \bm{X}^{d}_{-j})$ is independent of $X^{d}_i$, and hence $s_j$ is independent of $X^{d}_j$ and $X^{d}_i$.
By applying the same argument to all $i = 1, \ldots, K_d$, we can show that $s_i(Q, \bm{X}^{d}_{-i})$ is independent of all elements in $\bm{X}^{d}_{-i}$, that is, $s_i(Q, \bm{X}^{d}_{-i}) = s_i(Q)$ for all $i = 1, \ldots, K_d$.

By comparing \eqref{eq:inverse_demand_separable_step1} for all $i = 1, \ldots, K_d$, it is easy to see that $s_i(Q) = s_j(Q)$ for all $i,j = 1, \ldots, K_d$.
Therefore, we have a symmetry of $s_i$ for all $i = 1, \ldots, K_d$, that is, $s_i(Q) = s(Q)$ for all $i = 1, \ldots, K_d$.
Thus, $P(Q, \bm{X}^{d})$ must be of the form
\begin{align}
    P(Q, \bm{X}^{d}) = g(Q)r(\bm{X}^{d}) + s(Q). \label{eq:inverse_demand_separable_c_i_constant}
\end{align}
Note that when $g(Q) = 0$ for any $Q >0$, the inverse demand function depends only on the aggregate quantity, which violate Assumption \ref{assumption:effectiveness_shifters}.
Therefore, we assume that $g(Q) \ne 0$ for all $Q>0$.

\subsubsection*{Step 2: Necessity for $D_i$ is independent of $\bm{X}^{d}$}
Next, given the inverse demand function \eqref{eq:inverse_demand_separable_c_i_constant}, we further specify the form of the inverse demand function based on $D_i(Q, \bm{X}^{d})$.
By substituting \eqref{eq:inverse_demand_separable_c_i_constant} into \eqref{eq:interaction_derivative_demand}, we have
\begin{align}
    D_i(Q, \bm{X}^{d}) = &\frac{\theta^{*} - \theta}{r_i(\bm{X}^{d})(g(Q) + \theta Q g'(Q))}\Big[
    (g'(Q)r(\bm{X}^{d}) + s'(Q) )g(Q)r_i(\bm{X}^{d})\\
    & \hspace{4.5cm} + Q(g''(Q)r(\bm{X}^{d}) + s''(Q))g(Q)r_i(\bm{X}^{d}) \\
    &\hspace{4.8cm} - Q(g'(Q)r(\bm{X}^{d}) + s'(Q))g'(Q)r_i(\bm{X}^{d})\Big]\\
    =&\frac{\theta^{*} - \theta}{g(Q) + \theta Q g'(Q)}\Big[r(\bm{X}^{d})[g'(Q)g(Q) + Qg{''}(Q)g(Q) - Q (g'(Q))^{2}]\\
    & \hspace{3.5cm} + s'(Q)g(Q) + Qs''(Q)g(Q)- Qs'(Q)g'(Q)\Big].
\end{align}
The dependence of $D_i(Q, \bm{X}^{d})$ on $\bm{X}^{d}$ comes only from the first term in the bracket.
Therefore, $D_i(Q, \bm{X}^{d})$ is independent of $\bm{X}^{d}$ if and only if
\begin{align}
    \frac{\partial D_i}{\partial X^{d}_j}(Q, \bm{X}^{d}) =\frac{(\theta^{*} - \theta)r_j(\bm{X}^{d})}{g(Q) + \theta Q g'(Q)} [g'(Q)g(Q) + Qg{''}(Q)g(Q) - Q (g'(Q))^{2}] = 0.
\end{align}
As Assumption \ref{assumption:effectiveness_shifters} implies that $r_j(\bm{X}^{d})$ is nonzero, we must have the terms in the bracket to be zero:
\begin{align}
    g'(Q)g(Q) + Qg{''}(Q)g(Q) - Q (g'(Q))^{2} = 0. \label{eq:differential_equation_for_g}
\end{align}
As $g(Q) \ne 0$ for $Q >0$, let $v(Q) = \frac{g'(Q)}{g(Q)}$.
Because $v'(Q) = \frac{g''(Q)g(Q) - (g'(Q))^2}{g(Q)^2}$, dividing \eqref{eq:differential_equation_for_g} by $g(Q)^2$ gives
\begin{align}
    v(Q) + Qv'(Q) = 0. \label{eq:differential_equation_for_v}
\end{align}
This is a first-order linear differential equation.
To solve this differential equation, we consider two cases.

\paragraph{Case 1: $v(Q) = 0$ for all $Q>0$.}
This happens when $g(Q)$ is a constant.
In this case, \eqref{eq:differential_equation_for_g} holds immediately, and hence \eqref{eq:differential_equation_for_v} also holds.
Let $g(Q) = C$ for some nonzero constant $C$.
Then, we have
\begin{align}
    P(Q, \bm{X}^{d}) = Cr(\bm{X}^{d}) + s(Q).
\end{align}
For simplicity, we can absorb the constant $C$ into $r(\bm{X}^{d})$ and $s(Q)$, and hence we have
\begin{align}
    P(Q, \bm{X}^{d}) = r(\bm{X}^{d}) + s(Q). \label{eq:inverse_demand_separable_step2_constant}
\end{align}

\paragraph{Case 2: $v(Q) \ne 0$ for all $Q >0$.}
In this case, \eqref{eq:differential_equation_for_v} implies that
\begin{align}
    \frac{v'(Q)}{v(Q)} = -\frac{1}{Q}.
\end{align}
Then, it can be written as
\begin{align}
    \frac{d }{d Q}\log |v(Q)| = -\frac{d}{dQ}\log Q.
\end{align}
Integrating both sides with respect to $Q$, we have
\begin{align}
    \log |v(Q)| = -\log Q + a_1,
\end{align}
where $a_1 \in \mathbb{R}$ is a constant.
By taking the exponential of both sides, we have
\begin{align}
    |v(Q)| = \frac{\alpha}{Q}, 
\end{align}
where $\alpha = \exp(a_1)$ is a positive constant.
This has two solutions
\begin{align}
    v(Q) = \frac{g'(Q)}{g(Q)} = \pm \frac{\alpha}{Q}.
\end{align}
Since $\alpha > 0$, we redefine the constant as $\alpha \in \mathbb{R} \backslash \{0\}$ to absorb the sign into $\alpha$.
Then, we have 
\begin{align}
    \frac{d }{d Q}\log |g(Q)| = \alpha\frac{d}{dQ}\log Q.
\end{align}
Again, by integrating both sides with respect to $Q$, we have
\begin{align}
    \log |g(Q)| = \alpha\log Q + a_2,
\end{align}
where $a_2 \in \mathbb{R}$ is a constant.
By taking the exponential of both sides, we have
\begin{align}
    |g(Q)| = \alpha_2 Q^{\alpha},
\end{align}
where $\alpha_2 = \exp(a_2)$ is a positive constant.
Again, this has two solutions
\begin{align}
    g(Q) = \pm \alpha_2Q^{\alpha}.
\end{align}
Since $\alpha_2 > 0$, we can define $\alpha_2 \in \mathbb{R} \backslash \{0\}$ to absorb the sign into $\alpha_2$.
Therefore, when $C_i$ and $D_i$ are independent of $\bm{X}^{d}$, the inverse demand function must be of the form
\begin{align}
    P(Q, \bm{X}^{d}) = Q^{\alpha}r(\bm{X}^{d}) + s(Q). \label{eq:inverse_demand_separable_step2}
\end{align}
Here, $\alpha_2$ is absorbed into $r$.

While $\alpha$ is nonzero in \eqref{eq:inverse_demand_separable_step2}, by allowing that $\alpha = 0$, it can include \eqref{eq:inverse_demand_separable_step2_constant} as a special case.
At the same time, recall that we require that \eqref{eq:identification_condition_separable} does not hold for any $i$.
However, when $\alpha = -\frac{1}{\theta}$, we have the inverse demand function is equal to \eqref{eq:identification_separable_demand} where $\mathcal{I} = \emptyset$, which implies that the conduct parameter and the marginal cost function are identified. 
Therefore, we should have $\alpha \ne -\frac{1}{\theta}$ to ensure the non-identification of the conduct parameter.

\subsection{Proof of Lemma \ref{lemma:sufficient_nonidentification}}

From the proof of Lemma \ref{lemma:nonidentification_inverse_demand}, we know that $C_i$ and $D_i$ are independent of $\bm{X}^{d}$ under \eqref{eq:nonidentification_inverse_demand}.
Under the inverse demand function \eqref{eq:inverse_demand_separable_c_i_constant}, we have
\begin{align}
    \hat{C}_i(Q, \bm{X}^{d}) = \frac{Q^{\alpha}r_i(\bm{X}^{d}) + \theta^{*}\alpha Q^{\alpha}r_i(\bm{X}^{d})}{Q^{\alpha}r_i(\bm{X}^{d}) + \theta \alpha Q^{\alpha}r_i(\bm{X}^{d})} = \frac{1 + \theta^{*}\alpha}{1 + \theta\alpha}.
\end{align}
and
\begin{align}
    \hat{D}_i(Q, \bm{X}^{d}) & = \frac{\theta^{*} - \theta}{Q^{\alpha} + \theta \alpha Q^{\alpha}} \left[ s'(Q)Q^{\alpha} + Qs''(Q) Q^{\alpha} - s'(Q)\alpha Q^{\alpha} \right]\\
    &= \frac{\theta^{*} - \theta}{1 + \theta\alpha} \left[\frac{d}{dQ}Qs'(Q)  -\alpha s'(Q) \right].
\end{align}

Then, consider a transformation of the derivative of a function $f(Q, X)$ with respect to $Q$ based on \eqref{eq:mc_transformation_quantity} such that
\begin{align}
    T_Q\left(f(Q,X), Q\right) & \equiv \hat{D}_i(Q, \bm{X}^{d}) + \hat{C}_i(Q, \bm{X}^{d}) \frac{\partial f}{\partial Q}(Q, X)\\
    & = \frac{\theta^{*} - \theta}{1 + \theta\alpha} \left[\frac{d}{dQ}Qs'(Q)  -\alpha s'(Q) \right] + \frac{1 + \theta^{*}\alpha}{1 + \theta\alpha} \frac{\partial f}{\partial Q}(Q, X).
\end{align}
This can be integrated with respect to $Q$, and hence we obtain a transformation of $f(Q, X)$ as
\begin{align}
    T\left(f(Q,X), Q\right) \equiv \frac{\theta^{*} - \theta}{1 + \theta\alpha} \left[Qs'(Q) - \alpha s(Q) \right] + \frac{1 + \theta^{*}\alpha}{1 + \theta\alpha} f(Q, X).
\end{align}
Note that we assume the integral constant is zero for simplicity.

Define a marginal cost function $MC^{*}$ as
\begin{align}
    MC^{*}(Q, \bm{X}^{s}) \equiv \frac{\theta^{*} - \theta}{1 + \theta\alpha} \left[Qs'(Q) - \alpha s(Q) \right] + \frac{1 + \theta^{*}\alpha}{1 + \theta\alpha} MC(Q, \bm{X}^{s}).
\end{align}
Then, by substituting the marginal revenue under $\theta$ into $T$, we can obtain the marginal revenue under $\theta^{*}$:
\begin{align}
    & \frac{\theta^{*} - \theta}{1 + \theta\alpha} \left[Qs'(Q) - \alpha s(Q) \right] + \frac{1 + \theta^{*}\alpha}{1 + \theta\alpha} \left[(1+\theta\alpha) Q^{\alpha}r(\bm{X}^{d}) + s(Q) + \theta Qs'(Q)\right]\\
    = & (1 + \theta^{*}\alpha)Q^{\alpha}r(\bm{X}^{d}) + \frac{(\theta^{*} - \theta + (1 + \theta^{*}\alpha)\theta)Qs'(Q) + (1 + \theta^{*}\alpha - (\theta^{*} - \theta)\alpha) s(Q)}{1 + \theta\alpha}\\
    = & (1 + \theta^{*}\alpha)Q^{\alpha}r(\bm{X}^{d}) + \frac{(1 + \theta\alpha) s(Q) + \theta^{*}(1 + \theta\alpha)Qs'(Q) }{1 + \theta\alpha}\\
    = & (1 + \theta^{*}\alpha)Q^{\alpha}r(\bm{X}^{d}) +s(Q) + \theta^{*}Qs'(Q).
\end{align}
Then, suppose that $Q^e$ satisfies the equilibrium condition under $(\theta, MC)$.
Then, by using the definition of $MC^{*}$ and the above observation on the marginal revenue, we can confirm that $Q^e$ also satisfies the equilibrium condition under $(\theta^{*}, MC^{*})$:
\begin{align}
    & P(Q^e, \bm{X}^{d}) + \theta^{*} Q^e \frac{\partial P}{\partial Q}(Q^e, \bm{X}^{d}) - MC^{*}(Q^e, \bm{X}^{s})\\
    = & \frac{\theta^{*} - \theta}{1 + \theta\alpha} \left[Qs'(Q) - \alpha s(Q) \right] + \frac{1 + \theta^{*}\alpha}{1 + \theta\alpha} \left[(1+\theta\alpha) Q^{\alpha}r(\bm{X}^{d}) + s(Q) + \theta Qs'(Q)\right]\\
    & \hspace{1cm} - \frac{\theta^{*} - \theta}{1 + \theta\alpha} \left[Qs'(Q) - \alpha s(Q) \right] - \frac{1 + \theta^{*}\alpha}{1 + \theta\alpha} MC(Q^e, \bm{X}^{s})\\
    = & \frac{1 + \theta^{*}\alpha}{1 + \theta\alpha} \left[(1+\theta\alpha) Q^{\alpha}r(\bm{X}^{d}) + s(Q) + \theta^{*}Qs'(Q)-  MC(Q^e, \bm{X}^{s})\right] = 0.
\end{align}
The last equality holds because $Q^e$ satisfies the equilibrium condition under $(\theta, MC)$.
Therefore, the two models are observationally equivalent, which implies that the non-identification holds.

\section{Summary of Goldman and Uzawa (1964)}\label{appendix:summary_goldman_uzawa}

Let $n$ be the number of variables and $x = (x_{1},\ldots, x_{n})$ be a vector of $n$ variables.
Consider a partition of $X$ into $K$ parts, $\{x^1, \ldots, x^K\}$ such that $X = \bigcup_{k=1}^K x^k$ and $x^k \cap x^l = \emptyset$ for $k\ne l$.
\begin{definition}\label{def:weak_separable}
    A function is weakly separable with respect to the partition if 
    \begin{align}
        \frac{\partial}{\partial x_l}\left(\frac{\frac{\partial f}{\partial x_i}(x^1, \ldots, x^K)}{\frac{\partial f}{\partial x_j}(x^1, \ldots, x^K)}\right) = 0, \quad i,j\in x^k, l \notin x^k.
    \end{align}
\end{definition}
This implies that the ratio of the derivative with respect to $x_i$ and $x_j$, which are in the same category, is not affected by the change in the variables in other partitions.
Intuitively, by taking the ratio, the component of $f$ relating to $x_l$ is canceled out, and hence the derivative of the ratio with respect to $x_l$ becomes zero.
When $f$ is a utility function, this implies that the marginal rate of substitution between commodity $i$ and $j$ in the same partition is independent of the quantities of commodities outside $x^k$.

Then, \citet{goldmanNote1964} specifies the functional form that a weak separable function should satisfy.
\begin{theorem}\label{theorem_2_GU}
    A function $f(x)$ is weakly separable with respect to a partition $\{x^1, .. ., x^K\}$ if and only if $f(x)$ is of the form: 
    \begin{align}
        f(X) = \Phi(r^1(x^{1}),\ldots, r^K(x^{K})   )
    \end{align} where $\Phi(r^1,\ldots, r^K)$ is a function of $K$ variables and, for each $k$, $r^k(x^{k})$ is a function of subvector $x^{k}$ alone.
\end{theorem}

The next lemma is a key lemma in Lau's proof.
\begin{lemma}\label{lemma_1_GU}
    Let $f(x)$ and $g(x)$ be two continuously twice-differentiable functions of $n$ variables $x=(x_1, \dots, x_n)$. If each indifference surface is connected, and if there exists a function $\lambda(x)$ such that
    \begin{align}
    \frac{\partial f}{\partial x_i}(x) &= \lambda(x)\frac{\partial g}{\partial x_i}(x), \quad i=1, \dots, n, \quad \text{for all } x, \label{eq:transform_f}
    \end{align}
    then $f(x)$ is a transformation of $g(x)$; namely, there exists a function $T$ of one variable such that
    \begin{align}
    f(x) &= T(g(x)) \quad \text{for all } x.
    \end{align}
    Hence, in particular, the function $\lambda(x)$ satisfying \eqref{eq:transform_f} must be of the form:
    \begin{align}
        \lambda(x) &= \Lambda(g(x)) \quad \text{for all } x, \label{eq:form_of_lambda}
    \end{align}
    with some function $\Lambda$ of one variable.
\end{lemma}
If there is a function $T$ such that $f(x) = T(g(x))$, then by the chain rule, we have $\frac{\partial f}{\partial x_i}(x) = T'(g(x))\frac{\partial g}{\partial x_i}(x)$ where $T'$ is the derivative of $T$.
Thus, by defining $\lambda(x) = T'(g(x))$, we have $\frac{\partial f}{\partial x_i}(x) = \lambda(x)\frac{\partial g}{\partial x_i}(x)$.
Intuitively, the lemma implies that the converse of the chain rule holds under additional conditions.

\end{document}